\documentclass[12pt]{article}
\usepackage{epsfig,amssymb,array,pstricks,pst-coil}
\usepackage{amsmath}
\usepackage{graphics,psfrag,rotating,cite}

\def\partials{{\partial \mkern-9mu/}}
\def\lsim{\raisebox{-.4ex}{$\stackrel{<}{\scriptstyle \sim}$\,}}
\def\gsim{\raisebox{-.4ex}{$\stackrel{>}{\scriptstyle \sim}$\,}}

\makeatletter
\renewcommand{\section}{\setcounter{equation}{0}\@startsection
    {section}%
    {1}%
    {0pt}%
    {-1\baselineskip}%
    {0.4\baselineskip}%
    {\large\bfseries}}%
\renewcommand{\subsection}{\@startsection
    {subsection}%
    {2}%
    {0pt}%
    {-0.75\baselineskip}%
    {0.2\baselineskip}%
    {\bfseries}}%
\renewcommand{\subsubsection}{\@startsection
    {subsubsection}%
    {3}%
    {0pt}%
    {-0.5\baselineskip}%
    {0.1\baselineskip}%
    {\sc}}%
\makeatother

\begin{document}
\input epsf \renewcommand{\topfraction}{0.8}
\pagestyle{empty} \vspace*{-5mm}
\begin{center}
\Large{\bf Radiative corrections to the Higgs potential in the LH model}\\
\vspace*{2cm} \large{ Antonio Dobado, Lourdes Tabares-Cheluci}\\
\vspace{0.2cm} \normalsize
Departamento de  F\'{\i}sica Te\'orica I,\\
Universidad Complutense de
Madrid, E-28040 Madrid, Spain
\vspace{0.2cm}\\
\large{Siannah Pe\~naranda}
\\ \vspace{0.2cm} \normalsize
Departamento de F{\'{\i}}sica Te{\'o}rica, Universidad
de Zaragoza, E-50009, Zaragoza, Spain
\vspace{0.2cm}\\
\large{Javier Rodriguez-Laguna}
\\ \vspace{0.2cm} \normalsize
Departamento de Matem{\'a}ticas, Universidad Carlos III de Madrid,
E-28911, Madrid, Spain\\
\vspace*{2cm}{\bf ABSTRACT}
\end{center}
In this work we compute the radiative corrections to the Higgs
mass and the Higgs quartic couplings coming from the Higgs sector
itself and the scalar fields $\phi$ in the Littlest Higgs (LH)
model. The restrictions that the new contributions set on the
parameter space of the models are also discussed. Finally this
work, together with our three previous papers, complete our
program addressed to compute the relevant contributions to the
Higgs low-energy effective potential in the LH model and the
analysis of their phenomenological consequences.
 \noindent
\newpage
\setcounter{page}{1} \pagestyle{plain} \textheight 20 true cm

\section{Introduction}

The discovery of a Higgs boson and the elucidation of the
mechanism responsible for the electroweak symmetry breaking are
some of the major goals of present and future searches in particle
physics. Because of the precise data obtained for a long time
 to test the Standard Model (SM) of particle interactions, and the recent
measurements of the $W$ and the top masses at the Fermilab
Tevatron~\cite{mwmtmedida}, the SM has been confirmed as the right
model describing the electroweak phenomena at the current
experimental energy scale. However, the origin of the electroweak
symmetry breaking, for which the Higgs boson is responsible in the
SM, remains elusive. The quadratically divergent contributions to
the Higgs mass and the electroweak precision observables imply
different scales for physics beyond the SM, being the first one
below $1$ TeV and the second one above $10$ TeV. This is the so
called {\em{little hierarchy problem}}. As it is well known the
mass of the Higgs boson receives one-loop corrections that are
quadratic in the loop momenta. The largest contributions come from
the top quark loop, with smaller corrections coming from loops of
the electroweak gauge bosons and of the Higgs boson itself.
Cancellations between the top sector and other sectors must occur
in order to have the Higgs mass lighter than $200$ GeV as expected
from the electroweak precision test of the SM, which requires a
fine-tuning of one part in 100. As this situation is quite
unnatural various theories and models have been designed to solve
this problem.

An interesting attempt to deal with it is the so called
{\em{Littlest Higgs model}} (LH)~\cite{Cohen}, inspired in an old
suggestion by Georgi and Pais~\cite{Georgi}, which tries to solve
the little hierarchy problem by adding new particles with masses
\textit{O}(TeV) and symmetries which protect the Higgs mass from
those dangerous quadratically divergent contributions
(see~\cite{Schmaltz} and~\cite{review1} for reviews). These
particles include the Goldstone bosons (GB) corresponding to a
global spontaneous symmetry breaking (SSB) from the $SU(5)$ to the
$SO(5)$ group, a new third generation vector quark called $T$ and
the gauge bosons corresponding to an additional gauge group which
contains at least a $SU(2)_R$ and eventually a new hypercharge
$U(1)$. In this case, and contrary to the supersymmetric theories,
cancellation occurs between same-statistics particles. However, LH
models typically leave  uncanceled logarithmic divergencies which
requires additional new contributions at some higher scale to
preserve a small Higgs boson mass. Many of such models with
different {\it{theory space}} have been
constructed~\cite{Cohen,models}, and electroweak precision
constraints on various little Higgs models have been investigated
by performing global fits to the precision
data~\cite{Csaki1,Csaki2,LoganP,EWPO1,recentpheno}. The existence
of the different new states in these models could give rise to a
very rich phenomenology, which could be probed at the CERN Large
Hadron Collider (LHC)~\cite{Logan,Peskin}.

Nevertheless, it is clear that any viable model has to fulfill the
basic requirement of reproducing the SM model at low energies. In
particular, from the LH model it is possible in principle
 to compute the Higgs low-energy effective potential
and then, by comparing with the SM potential, to obtain their
phenomenological consequences including new restrictions on the
parameter space of the LH model itself. For example, one can
obtain the one-loop contribution to the parameters of the standard
Higgs potential,
\begin{equation}
\label{eq:SMpotential}
V=-\mu^{2}HH^{\dag}+\lambda (HH^{\dag})^{2};
\end{equation}
where $\mu^{2}$ and $\lambda$ denote the well known Higgs mass and
Higgs self-couplings parameters. Then it is possible to set
restrictions over the LH parameters space by imposing the
condition
 $\mu^2 = \lambda v^2$, where $v$ is the SM vacuum
expectation value ($H=(0,v)/\sqrt{2}$). The $\mu^{2}$ sign and
value are well known~\cite{Cohen,Peskin}, and effectively they are
the right ones to produce the electroweak symmetry breaking,
giving a Higgs mass $m_H^2=2 \mu^{2}$. However, the full
expression for the radiative corrections to $\lambda$ has not been
analyzed in detail so far. In principle both $\mu^2$ and $\lambda$
receive contributions from fermion, gauge boson and scalar loops,
besides others that could come from the ultraviolet completion of
the LH model. We have previously computed the contributions to the
Higgs effective potential in the LH model coming from the fermion
sector and the gauge boson sector~\cite{ATP,ATP2}. On the other
hand, several relations for the threshold corrections to the
$\lambda$ parameter in the presence of a $10$ TeV cut-off,
depending on the UV-completion of the theory, have been reported
(see, for example~\cite{italianos}). Besides, we have computed the
effective potential for the doublet Higgs and the triplet
$\phi$~\cite{ATP3}, coming from the fermionic and gauge boson
one-loop contributions and from the higher order effective
operators needed for the ultraviolet completion of the model.

In~\cite{ATP} and~\cite{ATP2} we computed and analyzed the fermion
contributions to the low energy Higgs effective potential together
with the effects of virtual heavy and electroweak gauge bosons
present in the LH model. We have illustrated in these works the
kind of constraints on the possible values of the LH parameters
that can be set by requiring the complete LH effective potential
to reproduce exactly the SM potential. The radiative corrections
to $\lambda$, at the one-loop level, had not been previously
computed. The computation of $\lambda$ is important for several
reasons: First, it must be positive, for the low energy effective
action to make sense. In addition, from the effective
potential~(\ref{eq:SMpotential}), one gets the simple formula
$m^2_H=2 \lambda v^2$ or, equivalently, $\mu^2 = \lambda v^2$,
where $v$ is set by the experiment (for instance from the muon
lifetime) to be $v\simeq 245$ GeV. In our phenomenological
discussion in~\cite{ATP,ATP2} we have shown that the one-loop
effective potential of the LH model cannot reproduce the SM
potential with a low enough Higgs mass, $m^2_H=2 \lambda v^2=2
\mu^2$, in agrement with the present experimental constraints.

In order to solve this problem we computed in~\cite{ATP3} the
effective potential for the doublet Higgs and the triplet $\phi$;
coming from the fermionic and gauge boson one-loop contributions
and also from the higher order effective operators, as defined
in~\cite{Logan}. The relevant terms of this effective potential
can be read as,
\begin{eqnarray}
\label{potef1}
V_{eff}(H,\phi)&=&-\mu_{fg}^{2}HH^{\dag}+\lambda_{fg}(HH^{\dag})^{2}\nonumber\\
&&+\lambda_{\phi^{2}} f^{2} \mbox{tr}(\phi \phi^{\dag})+
i\lambda_{H^2\phi}f(H\phi^{\dag}H^{T}-H^{*}\phi H^{\dag})\,,
\end{eqnarray}
where $\mu_{fg}^{2} > 0$ and $\lambda_{fg} > 0$.

With this potential we studied the regions of the LH parameter
space giving rise to the SM electroweak symmetry breaking.
Although radiative corrections from fermion and gauge boson loops
were discussed in~\cite{ATP,ATP2}, the radiative contributions to
$\lambda_{\phi^{2}}$ and $\lambda_{H^2\phi}$ have not been
computed so far. New constraints over the LH parameter space
emerge once we impose the new relation between coefficients of the
effective Higgs potential namely; $v^2={\mu_{fg}^{2}}/
{\lambda_{fg}-\lambda_{H^{2}\phi}^{2}/\lambda_{\phi^{2}}}$. In
particular, the lowest value found for the $\mu$ parameter was
$390$ GeV~\cite{ATP3}, which implied  a Higgs boson mass of about
$m_{H}\simeq 550$ GeV,  still not compatible with the present
experimental constraints.

On the other hand it is well known that the radiative corrections
coming from the Higgs itself and the $\phi$ fields could also
provide relevant contributions to the effective potential. Thus
the main goal of the present work is to check wether these
corrections could really reduce the Higgs mass to solve the above
mentioned problem, making the LH model compatible with the present
phenomenology.

This work is organized as follows: In Section 2 we briefly explain
the LH model. A summary on the SSB and the mass eigenstates is
presented in Section 3. We set the notation in the two aforementioned
sections. Section 4 is devoted to the computation of the radiative
corrections contributions to the Higgs mass and quartic coupling
coming from the scalar sector loops. In Section 5 we analyze the
constraints that our computation establishes on the LH parameters
and, finally, in Section 6 we present the conclusions. The
expressions of the coefficients of the effective potential
(\ref{potef1}) coming from the radiative corrections and the
effective operators are listed in the Appendix.

\section{The model}

The LH model is based on the assumption that there is a physical system with a global $SU(5)$
symmetry that is spontaneously broken to a $SO(5)$ symmetry at a high
scale $\Lambda$ through a vacuum expectation value ({\it{v.e.v}}) of
order $f$.  Thus, 14 Goldstone bosons (GB) are obtained as a
consequence of this breaking. In this work we will consider two
different versions of the LH model. In the first one the $SU(5)$
subgroup $[SU(2)\times U(1)]^2$ is gauged. We refer to this version as
\emph{Model I}. In the second one the gauge group is $[SU(2)^{2}\times
U(1)]$ (\emph{Model II}) ~\cite{ATP,ATP2}. In both cases some of the
GB acquire masses through radiative corrections coming from the gauge
bosons and the $t$, $b$ and $T$ fermions loops.

The starting Lagrangian of the LH model is given
by~\cite{Cohen,Logan,Peskin}:
\begin{equation}\label{Ltotal}
\textit{L}=\textit{L}_{\Sigma}+\textit{L}_{YK}
\end{equation}
where $\textit{L}_{\Sigma}$ is the Non Linear Sigma Model  (NLSM)
lagrangian:
\begin{equation}\label{Lsigma}
\textit{L}_{\Sigma}=\frac{f^2}{8}\mbox{tr} [(D_{\mu}\Sigma)
(D^{\mu}\Sigma)^\dag]\,;
\end{equation}
and $\textit{L}_{YK}$ the Yukawa couplings for fermions and
scalars:
\begin{equation}\label{Lyukawa}
\textit{L}_{YK}=-\frac{\lambda_{1}}{2}f
\overline{u}_{R}\epsilon_{mn}\epsilon_{ijk}\Sigma_{im}\Sigma_{jn}\chi_{Lk}
-\lambda_{2} f \overline{U}_{R}U_{L}+\mbox{h.c.}\,.
\end{equation}
In the above Lagrangians $\Sigma$ is the GB matrix given by:
\begin{equation}
\Sigma=e^{2 i\Pi/f} \Sigma_{0}
\end{equation}
where $\Sigma_{0}$ can be chosen to be:
\begin{equation}
\Sigma_{0}= \left(%
\begin{array}{ccc}
  0 & 0 & \textbf{1} \\
  0 & 1 & 0 \\
  \textbf{1} & 0 & 0 \\
\end{array}%
\right)\,,
\end{equation}
with  $\textbf{1}$ being the $2 \times 2$ unit matrix, and the
$\Pi$ matrix can be parameterized as:
\begin{eqnarray}
\Pi & = & \left(%
\begin{array}{ccc}
0& \frac{-i}{\sqrt{2}}H^{\dag} & \phi^{\dag} \\
  \frac{i}{\sqrt{2}}H & 0 & \frac{-i}{\sqrt{2}}H^{*} \\
  \phi & \frac{i}{\sqrt{2}}H^{T} & 0 \\
\end{array}%
\right),
\end{eqnarray}
where$H=(H^{0},H^{+})$ is the SM Higgs doublet and $\phi$ is the
triplet given by:
\begin{equation}
\phi=\left(%
\begin{array}{cc}
  \phi^{0} & \frac{1}{\sqrt{2}}\phi^{+} \\
\frac{1}{\sqrt{2}}\phi^{+} & \phi^{++}
\end{array}%
\right)\,.
\end{equation}
The covariant derivative $D_{\mu}$ is defined as:
\begin{eqnarray}
\mbox{\emph{Model I}}&&\nonumber\\
D_{\mu}\Sigma & = &
\partial_{\mu}\Sigma-i\sum_{k=1}^{2}g_kW^a_k(Q_k^a\Sigma +\Sigma
Q_k^{aT})- i\sum_{k=1}^{2}g'_kB_k(Y_k\Sigma+\Sigma Y_k^{T})\nonumber\\
\mbox{\emph{Model II}}&&\nonumber\\
D_{\mu}\Sigma & = &
\partial_{\mu}\Sigma-i\sum_{k=1}^{2}g_kW^a_k(Q_k^a\Sigma +\Sigma
Q_k^{aT})- i g'B(Y\Sigma+\Sigma Y^{T})\,,
\end{eqnarray}
where $g$ and $g'$ are the gauge couplings, $W_{k}^{a}$
$(a=1,2,3)$ and $B_{k}\,, B$ are the $SU(2)$ and $U(1)$ gauge
fields respectively,  $Q_{1ij}^a=\sigma_{ij}^a/2$  for $i,j=1,2$
and zero otherwise, $Q_{2ij}^{a}=\sigma_{i-3,j-3}^{a*}/2$ for
$i,j=4,5$ and zero otherwise, $Y_{1}= diag(-3,-3,2,2,2)/10$,
$Y_{2}= diag(-2,-2,-2,3,3)/10$  and $Y= diag(-1,-1,0,1,1)/2$. The
Yukawa Lagrangian in~(\ref{Lyukawa}) describes the interactions
between GB and  fermions, more exactly, the third generations of
quarks plus the extra $T$ quark appearing in the LH model. The
indices in $\textit{L}_{YK}$ are defined such that $m,n=4,5$,
$i,j=1,2,3$, and
\begin{eqnarray}
\overline{u}_{R}&=& c \,\overline{t}_{R}+ s  \,\overline{T}_{R}\,,\nonumber\\
\overline{U}_{R}&=&-s \,\overline{t}_{R}+   c \,\overline{T}_{R},
\end{eqnarray}
with:
\begin{eqnarray}
c&=&\cos
\theta=\frac{\lambda_{2}}{\sqrt{\lambda_{1}^{2}+\lambda_{2}^{2}}},\nonumber\\
s&=&\sin \theta =
\frac{\lambda_{1}}{\sqrt{\lambda_{1}^{2}+\lambda_{2}^{2}}}\,,
\end{eqnarray}
and
\begin{equation}
\chi_{L}=\left(%
\begin{array}{c}
  u \\
  b \\
  U \\
\end{array}%
\right)_{L}=\left(%
\begin{array}{c}
  t \\
  b \\
  T \\
\end{array}%
\right)_{L}.
\end{equation}
In addition to the above terms it is needed to add to the LH
Lagrangian the Yang-Mills terms corresponding to the various gauge
fields, and also the gauge fixing and Faddeev-Popov terms. Some of
the gauge fields get massive at the tree level through the Higgs
mechanism associated to the $SU(5)/SO(5)$ symmetry breaking. By
using the Landau gauge, which is the most appropriate for the kind
of computations we are presenting here (see \cite{ATP2} for
further details), the quadratic part of the complete gauge boson
Lagrangian can be written as:
\begin{equation}\label{Lomega}
\textit{L}_{\Omega}=\frac{1}{2}\Omega^{\mu}((\Box+M_{\Omega}^{2})g_{\mu\nu}
-\partial_{\mu}\partial_{\nu}+2 \tilde{I} \,g_{\mu\nu})\Omega^{\nu}\,,
\end{equation}
where $\Omega$ stands for any of the gauge bosons,
\begin{eqnarray}
\mbox{\emph{Model I}} \hspace{0.5cm} &&
\Omega^{\mu}=({W'}^{\mu a},W^{\mu a},{B'}^{\mu},B^{\mu}), \nonumber\\
\mbox{\emph{Model II}} \hspace{0.5cm} &&
\Omega^{\mu}=({W'}^{\mu a},W^{\mu a},B^{\mu}) \,,
\end{eqnarray}
being the mass matrix eigenstates,
\begin{eqnarray}
\mbox{\emph{Model I}} \hspace{0.5cm} &&
M_{\Omega}=(M_{W'} 1_{3\times 3},0_{3\times 3},M_{B'},0),\nonumber\\
\mbox{\emph{Model II}} \hspace{0.5cm} &&
M_{\Omega}=(M_{W'} 1_{3\times 3},0_{3\times 3},0)\,,
\end{eqnarray}
with $M_{W'}= f \sqrt{g_1^2+g_2^2}/2$ and $M_{B'}= f
\sqrt{g_1^{'2}+g_2^{'2}}/\sqrt{20}$. The gauge boson mass
eigenstates are defined such as:
\begin{eqnarray}
W^a   & = & c_{\psi} W_1^a+ s_{\psi} W_2^a, \nonumber\\
 W^{'a}  & = & s_{\psi} W_1^a- c_{\psi} W_2^a,
\end{eqnarray}
where
\begin{eqnarray}
s_{\psi} &=& \sin \psi = \frac{g_1}{\sqrt{g_1^2+g_2^2}}, \nonumber\\
 c_{\psi} &=& \cos \psi =
\frac{g_2}{\sqrt{g_1^2+g_2^2}},
\end{eqnarray}
and
\begin{eqnarray}
B  & = & c'_{\psi} B_1+ s'_{\psi} B_2, \nonumber\\
B'  & = & s'_{\psi} B_1- c'_{\psi} B_2,
\end{eqnarray}
with
\begin{eqnarray}
s'_{\psi}=  \sin \psi' = \frac{g'_{1}}{\sqrt{{g'}_{1}^{\,2}+{g'}_{2}^{\,2}}},
\nonumber\\
 c'_{\psi}= \cos \psi' =
\frac{{g'}_{2}}{\sqrt{{g'}_{1}^{\,2}+{g'}_{2}^{\,2}}}\,.
\end{eqnarray}
$\tilde{I}$ is the interaction matrix between the gauge bosons and
the $H$ and $\phi$ scalars which can be found in our previous
works ~\cite{ATP2,ATP3}.

By adding the appropriate kinetic terms, the complete Lagrangian
for the quarks becomes:
\begin{eqnarray}
\textit{L}_{\chi}= \overline{\chi}_{R}(i
\partials-M+\hat{I}) \chi_{L}+\mbox{h.c.}\,,
\end{eqnarray}
where
$$\chi_{R}=\left(%
\begin{array}{c}
  t \\
  b \\
  T \\
\end{array}%
\right)_{R}\,,$$ $M=$diag$(0,0,m_T)$ with
$m_{T}=f\sqrt{\lambda_{1}^{2}+\lambda_{2}^{2}}$ and $\hat{I}$ is
the scalar-quark interaction matrix. The elements of this matrix
can be found in ~\cite{ATP,ATP3}. For more details about the
model, including Feynman rules and also some phenomenological
results see for example~\cite{Logan}.

\section{Effective operators}
\label{sec:effoperator}

It is well known that the effective Higgs potential receive also
contributions from additional operators coming from the
ultraviolet completion of the LH model. Obviously these operators
must be consistent with the symmetries of the LH model
~\cite{Cohen,Logan,Casas}. At the lowest order they can be
parameterized by two unknown coefficients $a$ and $a'$ $\sim
O(1)$. The form of these effective operators is, for the fermion
sector~\cite{Logan}:
\begin{equation}
\textit{O}_{f}=-a'\frac{1}{4}\lambda_{1}^{2}f^{4}\epsilon^{wx}\epsilon_{yz}\epsilon^{ijk}\epsilon_{kmn}\Sigma_{iw}\Sigma_{jx}\Sigma^{*my}\Sigma^{*nz}\,,
\end{equation}
where $i,j,k,m,n$ run over 1,2,3 and $w,x,y,z$ run over 4,5 and
for
 the gauge sector  we have for \emph{Model I}:
\begin{eqnarray}
\textit{O}_{gb}=\frac{1}{2}af^{4}\left\{g_{j}^{2}\sum_{a=1}^{3}\mbox{Tr}\left[(Q_{j}^{a}\Sigma)(Q_{j}^{a}\Sigma)^{*}\right]+g_{j}^{'2}\mbox{Tr}\left[(Y_{j}\Sigma)(Y_{j}\Sigma)^{*}\right]\right\}\,,
\end{eqnarray}
with $j=1,2$ and $Q_{j}^{a}$ and  $Y_{j}$ being the generators of
the $SU(2)_{j}$ and $U(1)_{j}$ groups, respectively. In the case
of  \emph{Model II}:
\begin{eqnarray}
\textit{O}_{gb}=\frac{1}{2}cf^{4}\left\{g_{j}^{2}\sum_{a=1}^{3}\mbox{Tr}\left[(Q_{j}^{a}\Sigma)(Q_{j}^{a}\Sigma)^{*}\right]+g^{'2}\mbox{Tr}\left[(Y\Sigma)(Y\Sigma)^{*}\right]\right\}\,,
\end{eqnarray}
where $j=1,2$ and $Y$ is the generator of the unique $U(1)$ group.

By expanding the GB field matrix $\Sigma$ in these effective
operators, we obtain their different contributions to the
coefficients of the effective potential (\ref{potef1}). The
results are presented in the Appendix.

The complete result for the coefficients of the Higgs potential is
given by the sum of the contributions coming from the effective
operators, as given above, and the radiative contributions coming
from all sectors of the model, as will be discussed in the
following.

\section{SSB and mass eigenstates}

In the LH model the electroweak symmetry breaking is triggered, in
principle, by the Higgs potential generated by one-loop radiative
corrections, including both, fermion and gauge boson loops, and
the effective operators introduced in the previous section.
Obviously, this potential is invariant under the electroweak gauge
group $SU(2)\times U(1)$ and also should have the correct form to
break this symmetry spontaneously to $U(1)_{em}$. The relevant
terms for this work are given in~(\ref{potef1}). Quartic terms
involving $\phi^4$ and $H^2 \phi^2$ are not included since they
give subleading contributions to the Higgs mass. These parameters
were computed in our previous works \cite{ATP,ATP2,ATP3} and  are
given in the Appendix for completeness.

The scalar potential, as given in ~(\ref{potef1}), reaches its
minimum at:

$\langle HH^{\dag}\rangle = {v^2}/{2}$ and $\,\,\langle
\phi\phi^{\dag}\rangle = {v'}^2$ with:
\begin{equation} \label{minimun}
v^2=\frac{\mu_{fg}^{2}}{\lambda_{fg}-\lambda_{H^{2}\phi}^{2}/\lambda_{\phi^{2}}},  \hspace{2cm}
{v'}^2=\frac{\lambda_{H^{2}\phi}}{\sqrt{2}\lambda_{\phi^{2}}}\frac{v^2}{f}.
\end{equation}
Note that both, the doublet and triplet scalars, get a
{\it{v.e.v.}}, $v$ and $v'$ respectively. A standard choice for
the components of these fields at the vacuum is:
\begin{eqnarray}
H^{+}=0, \hspace{1cm} H_{0}=\frac{v}{\sqrt{2}},
\hspace{1cm} \phi_{0}= -v', \hspace{1cm} \phi^{+}=\phi^{++}=0.
\end{eqnarray}
Then $H$ and $\phi$ can be parameterized as:
\begin{eqnarray}
H =(w^{+},\frac{1}{\sqrt{2}}(v+h+iw_{0}) )
\hspace{0.3cm} \mbox{and} \hspace{0.3cm}
\phi=\left(%
\begin{array}{cc}
  -v'+\frac{1}{\sqrt{2}}(\xi+i\rho) & \frac{1}{\sqrt{2}}\phi^{+} \\
\frac{1}{\sqrt{2}}\phi^{+} & \phi^{++}
\end{array}%
\right).
\end{eqnarray}
Obviously the new fields describe fluctuations around the
 vacuum and the potential written in terms of them can be split
in four sectors, namely, the scalar, the pseudoscalar, the charged
 and the doubly charged. For the first three sectors we find that
the new fields are not mass eigenstates. By diagonalizing the
corresponding mass matrices we obtain the mass eigenstates in each
case. I.e., for the scalar sector:
\begin{eqnarray}\label{ss}
h &=& c_{0}\mathcal{H} + s_{0}\Phi_{0},\hspace{2cm} m^2_{\mathcal{H}}\equiv m^{2}_{fg} = 2\,\mu_{fg}^{2}, \nonumber\\
\xi &=& c_{0}\Phi_{0} - s_{0}\mathcal{H}, \hspace{2cm}
m^{2}_{\Phi_{0}} = M_{\phi}^{2}+2\, m^{2},
\end{eqnarray}
the pseudoscalar sector:
\begin{eqnarray}\label{ps}
w_{0} &=& c_{P}G^{0} + s_{P}\Phi^{P},\hspace{2cm} m^{2}_{G^{0}} = 0, \nonumber\\
\rho &=& c_{P}\Phi^{P} - s_{P}G^{0}, \hspace{2cm} m^{2}_{\Phi^{P}}
= M_{\phi}^{2}+2 m^{2},
\end{eqnarray}
and the charged sector:
\begin{eqnarray} \label{cs}
w^{+} &=& c_{+}G^{+} + s_{+}\Phi^{+}, \hspace{1.5cm}   m^{2}_{G^{+}} = 0, \nonumber\\
\phi^{+} &=& c_{+}\Phi^{+} + s_{+}G^{+}, \hspace{1.5cm}   m^{2}_{\Phi^{+}} = M_{\phi}^{2}+ m^{2},
\end{eqnarray}
with $M_{\phi}^2=\lambda_{\phi^{2}}f^{2}$,
$m^{2}=v^{2}\lambda_{H^{2} \phi}^{2}/\lambda_{\phi^{2}}$. The
doubly charged sector remains unchanged with a mass $M_{\phi}$.

 Where the  notation introduced for the mass
eigenstates is the following: $\mathcal{H}$ and $\Phi_{0}$ are
neutral scalars, $\Phi^{P}$ is a neutral pseudoscalar, $\Phi^{+}$
and $\Phi^{++}$ are the charged and doubly charged scalars, and
$G^{+}$ and $G^{0}$ are the would-be Goldstone bosons
corresponding  to the SM $W$ and $Z$.

In terms of the mass eigenstates the leading order in the
$\mathcal{O}(v^{2}/f^{2})$ expansion of the potential is given by:
\begin{eqnarray}\label{potef2}
V_{eff} &=& \frac{1}{2} m_{fg}^{2}
\mathcal{H}^{2}+\frac{1}{2}m_{\Phi_{0}}^{2}\Phi_{0}^{2}+
\frac{1}{2}m_{\Phi^{p}}^{2}{\Phi^{P}}^{2}\nonumber \\
&+& m_{\Phi^{+}}^{2}\Phi^{+}\Phi^{-}+v\lambda_{fg}\mathcal{H}^{3}
+v\lambda_{fg}{G^{0}}^{2}\mathcal{H}+2v\lambda_{fg}{G^{+}}{G^{-}}\mathcal{H}\nonumber \\
&+&\frac{\lambda_{fg}}{4}\mathcal{H}^{4}+\frac{\lambda_{fg}}{2}\mathcal{H}^2{G^{0}}^{2}
+\lambda_{fg}\mathcal{H}^2{G^{+}}{G^{-}}\nonumber \\
&-&\frac{\lambda_{H^{2}\phi}}{\sqrt{2}}f\mathcal{H}^2\Phi_{0}-\sqrt{2}\lambda_{H^{2}\phi}f\mathcal{H}{G^{0}}\Phi^{P}-\lambda_{H^{2}\phi}f(\mathcal{H}{G^{-}}\Phi^{+}+\mathcal{H}{G^{+}}\Phi^{-})+ ... \nonumber \\
\end{eqnarray}

\section{Goldstone boson sector contributions}

The objective of this section is the computation of the radiative
contributions to the Higgs mass and the Higgs quartic coupling
coming from the GB sector. The relevant Lagrangian is given by:
\begin{eqnarray}
\textit{L}_{GB}&=&\frac{1}{2}(\partial_{\mu}\Pi)(\partial^{\mu}\Pi)+\frac{1}{f^2}
\left((\partial_{\mu}\Pi)(\partial^{\mu}\Pi)\Pi\Pi+
\Pi(\partial_{\mu}\Pi)\Pi(\partial^{\mu}\Pi)\right)-V_{eff}.\nonumber\\
\end{eqnarray}

In order to calculate the radiative contributions we write this
Lagrangian in terms of the mass eigenstates and we split the Higgs
field as $\mathcal{H}=\mathcal{\overline{H}}+\mathcal{\tilde {H}}$
where $\mathcal{\overline{H}}$   is the vacuum field and
$\mathcal{\tilde {H}}$ describes the field fluctuations around
this point. Then the first two terms of the Lagrangian above
become:
\begin{eqnarray}
\textit{L}_{Kin}&=&\frac{1}{2}\left(1+2\frac{\mathcal{\overline{H}}^{2}}{f^2}\right)(\partial_{\mu}\mathcal{\tilde{H}})(\partial^{\mu}\mathcal{\tilde{H}})+
\frac{1}{2}\left(1+\frac{\mathcal{\overline{H}}^{2}}{2f^2}\right)(\partial_{\mu}\Phi_{0})(\partial^{\mu}\Phi_{0})\nonumber \\
&+&\frac{1}{2}\left(1+\frac{\mathcal{\overline{H}}^{2}}{2f^2}\right)(\partial_{\mu}{G^{0}})(\partial^{\mu}{G^{0}})
+\frac{1}{2}\left(1+\frac{\mathcal{\overline{H}}^{2}}{2f^2}\right)(\partial_{\mu}\Phi^{P})(\partial^{\mu}\Phi^{P})\nonumber \\
&+&\left(1+\frac{\mathcal{\overline{H}}^{2}}{4f^2}\right)(\partial_{\mu}\Phi^{+})(\partial^{\mu}\Phi^{-})
+\left(1+\frac{\mathcal{\overline{H}}^{2}}{2f^2}\right)(\partial_{\mu}{G^{+}})(\partial^{\mu}{G^{-}})\nonumber\\
&+&(\partial_{\mu}\Phi^{++})(\partial^{\mu}\Phi^{--}).
\end{eqnarray}
Obviously, all the kinetic terms in this formula, but the last
one, are not properly normalized. Therefore we write the fields in
terms of a new set of properly normalized fields up to order
$1/f^2$ as:
\begin{eqnarray}
\Upsilon &=& \left(1-\frac{\mathcal{\overline{H}}^{2}}{4f^{2}}\right) \Upsilon' \hspace{1cm} \mbox{with} \hspace{1cm} \Upsilon^{(')} = {G^{0(')}}, G^{\pm (')}, \Phi_{0}^{(')}, \Phi^{P(')},  \\
\mathcal{\tilde{H}} &=& \left(1-\frac{\mathcal{\overline{H}}^{2}}{f^{2}}\right)\mathcal{H}', \\
\Phi^{\pm} &=&
\left(1-\frac{\mathcal{\overline{H}}^{2}}{8f^{2}}\right)
\Phi^{\pm'},
\end{eqnarray}
so that the Lagrangian is just:
\begin{eqnarray}
\textit{L}_{Kin}&=&\frac{1}{2}\partial_{\mu}\mathcal{H}'\partial^{\mu} \mathcal{H}'+\frac{1}{2}\partial_{\mu}\Phi_{0}'\partial^{\mu} \Phi_{0}'+\frac{1}{2}\partial_{\mu}{G^{0}}'\partial^{\mu}{G^{0}}'+ \frac{1}{2}\partial_{\mu}{\Phi^{P}}'\partial^{\mu}{\Phi^{P}}'\nonumber\\
&+&\partial_{\mu}G^{+'}\partial^{\mu}G^{-'}+\partial_{\mu}\Phi^{+'}\partial^{\mu}\Phi^{-'}
+\partial_{\mu}\Phi^{++}\partial^{\mu}\Phi^{--},
\end{eqnarray}

Then,  the effective potential $V_{eff}$ is given by:
\begin{eqnarray}
V_{eff}= \frac{1}{2}m_{fg}^{2}\mathcal{\overline{H}}^2+\frac{\lambda_{fg}}{4}\mathcal{\overline{H}}^4+ V_{eff}^{ss}+V_{eff}^{ps}+V_{eff}^{cs}+...
\end{eqnarray}
where
\begin{eqnarray}
\label{Vss}
V_{eff}^{ss}&=&\frac{1}{2}m_{\Phi_{0}}^{2}\Phi_{0}^{'2}+\frac{1}{2}m_{fg}^{2}\mathcal{H}^{'2}+\frac{3}{2}\lambda_{fg}\overline{\mathcal{H}}^{2}\mathcal{H}^{'2}-\frac{\lambda_{\phi^{2}}}{4}\overline{\mathcal{H}}^{2}\Phi_{0}^{'2}\nonumber \\
&-&\sqrt{2}\lambda_{H^{2}\phi}f\overline{\mathcal{H}}\mathcal{H}'\Phi_{0}'-\frac{\lambda_{H^{2}\phi}}{\sqrt{2}}f\overline{\mathcal{H}}^{2'}\Phi_{0}',\\
\label{Vps}
V_{eff}^{ps}&=&\frac{1}{2}m_{\Phi^{p}}^{2}{\Phi^{P}}^{'2}+\frac{\lambda_{fg}}{2}\overline{\mathcal{H}}^{2}{G^{0}}^{'2}-\frac{\lambda_{\phi^{2}}}{4}
\overline{\mathcal{H}}^{2} {\Phi^{P}}^{'2}\nonumber \\ &-&\sqrt{2}\lambda_{H^{2}\phi}f\overline{\mathcal{H}}{G^{0}}'{\Phi^{P}}',\\
\label{Vcs}
V_{eff}^{cs}&=&m_{\Phi^{+}}^{2}\Phi^{'+}\Phi^{'-}+\lambda_{fg}\overline{\mathcal{H}}^{2}G^{'+}G^{'-}-\frac{\lambda_{\phi^{2}}}{4}\overline{\mathcal{H}}^{2}\Phi^{'+}\Phi^{'-}\nonumber\\
&-&\lambda_{H^{2}\phi}f\overline{\mathcal{H}}(G^{'-}\Phi^{'+}+G^{'+}\Phi^{'-}),
\end{eqnarray}

Observe that the third terms in~(\ref{Vss}), (\ref{Vps}) and
(\ref{Vcs}) describe the new interactions which come from the new
normalization of the fields and the fact that
the triplet boson mass is $\mathcal{O}(f^{2})$. These interactions
play a decisive role to cancel the quadratic divergences that come
from the GB loops. 

Finally, we can see that the split into different scalar sectors
is maintained after diagonalization and normalization. This fact
is important in order to simplify the computation of the radiative
contributions coming from the GB. Thus we can deal with each
scalar sector in an independent way  being the computations in all
cases similar. We illustrate this by computing the ($\mathcal{H'},
\Phi_0'$) contribution and then we apply the same method  to the
other scalars.

\subsection{Scalar sector contribution}

The Lagrangian for the scalar sector  ($\mathcal{H'}, \Phi_0'$) is
given by:
\begin {eqnarray} \label{Lss}
\mathcal{L}^{ss}(\overline{\mathcal{H}},\mathcal{H}',\Phi_{0}')&=&\frac{1}{2}\partial_{\mu}\mathcal{H}'\partial^{\mu}\mathcal{H}'+\frac{1}{2}\partial_{\mu}\Phi_{0}'\partial^{\mu}\Phi_{0}'- V_{eff}^{ss} \nonumber \\
&=& \frac{1}{2}\partial_{\mu}\mathcal{H}'\partial^{\mu}\mathcal{H}'+\frac{1}{2}\partial_{\mu}\Phi_{0}'\partial^{\mu}\Phi_{0}'-\frac{1}{2}m_{fg}^{2}\mathcal{H}^{'2}-\frac{1}{2}m_{\Phi_{0}}^{2}\Phi_{0}^{'2} \nonumber \\
&-&\frac{3}{2}\lambda_{fg}\overline{\mathcal{H}}^{2}\mathcal{H}^{'2}+\frac{\lambda_{\phi^{2}}}{4}\overline{\mathcal{H}}^{2}\Phi_{0}^{'2}+\sqrt{2}f\lambda_{H^{2}\phi}\overline{\mathcal{H}}\mathcal{H}'\Phi_{0}'+\frac{\lambda_{H^{2}\phi}}{\sqrt{2}}\overline{\mathcal{H}}^{2}\Phi_{0}'.\nonumber \\
\end {eqnarray}
The effective action for the $\overline{\mathcal{H}}$ is:
\begin{equation}
e^{iS_{eff}[\overline{\mathcal{H}}]}=\int
[d\mathcal{H}'][d\Phi_{0}'] e^{i\int dx \mathcal{L}^{ss}},
\end{equation}

From the (\ref{Lss}) we observe that the integration can be
computed in two steps: First we concentrate on the $\Phi_{0}'$
field and then we integrate the $\mathcal{H'}$ field. After
integrating $\Phi_{0}'$ we get the $\mathcal{H}$ effective action:
\begin{eqnarray} \label{Sphi0}
S_{eff}^{ss}[\overline{\mathcal{H}},\mathcal{H'}]&=&-\frac{i}{2}\mbox{Tr}\log\left[1+G_{\Phi_{0}}\frac{\lambda_{\phi^{2}}}{2}\overline{\mathcal{H}}^2\right]\nonumber \\
&-&f^{2}\lambda_{H^{2}\phi}^{2}\int dx dy \overline{\mathcal{H}}^{2} \mathcal{H}'_{x}G_{\Phi_{0}xy}\mathcal{H}'_{y}-\frac{\lambda_{H^{2}\phi}^{2}}{4}f^{2}\int dx dy G_{\Phi_{0}xy}\overline{\mathcal{H}}^{4}\delta_{yx}\nonumber\\
&=&-\frac{i}{2}\sum_{k=1}^{\infty}\frac{(-1)^{k+1}}{k}\mbox{Tr}\left(G_{\Phi_{0}}\frac{\lambda_{\phi^{2}}}{2}\overline{\mathcal{H}}^{2}\right)^{k}+\tilde{I}_{2}+\tilde{I}_{4}\,,
\end{eqnarray}
 where the $\Phi_{0}'$ propagator is given by:
\begin{equation}
G_{\Phi_{0}}(x,y)=\int d\tilde{k}e^{ik(x-y)}\frac{1}{k^{2}-m_{\Phi_{0}}^2},
\end{equation}
here $\tilde{k}\equiv d^{4}k/(2\pi)^4$,
and
\begin{eqnarray}
\tilde{I}_{2}&=&-f^{2}\lambda_{H^{2}\phi}^{2}\int dx dy \overline{\mathcal{H}}^{2} \mathcal{H}'_{x}G_{\Phi_{0}xy}\mathcal{H}'_{y}\,, \\
\tilde{I}_{4}&=&-\frac{\lambda_{H^{2}\phi}^{2}}{4}f^{2}\int dx dy \overline{\mathcal{H}}^{4}\delta_{xy}G_{\Phi_{0}xy}\,.
\end{eqnarray}

Observe that we have obtained three terms. The first and the third
ones are $\mathcal{H}'$ independent
 and they will give the $\Phi_{0}'$ radiative contributions to the Higgs mass and the quartic coupling.

Now integrating out $\mathcal{H}'$ we find its contribution to the
$\overline{\mathcal{H}}$ effective action:
\begin{eqnarray}\label{SH}
S^{ss}[\overline{\mathcal{H}}]&=&-\frac{i}{2}\mbox{Tr}\log\left[1+G_{\mathcal{H}'}(-3\lambda_{fg}
\overline{\mathcal{H}}^{2}+2\tilde{I}_{2})\right]\nonumber\\
&=&
\frac{i}{2}\sum_{k=1}^{\infty}\frac{(-1)^{k+1}}{k}\mbox{Tr}\left(G_{\mathcal{H}}(3\lambda_{fg}
\overline{\mathcal{H}}^{2}-2\tilde{I}_{2})\right)+...\,,
\end{eqnarray}
where $G_{\mathcal{H}'}$ is the $\mathcal{H}'$ propagator,
\begin{equation}
G_{\mathcal{H}'}(x,y)=\int d\tilde{k}e^{ik(x-y)}\frac{1}{k^{2}-m_{fg}^{2}}\,.
\end{equation}

Finally, by taking into account (\ref{Sphi0}) and (\ref{SH}), we
obtain the  $\overline{{\mathcal{H}}}$ effective action which
reads:
\begin{eqnarray}\label{Sss}
S^{ss}[\overline{\mathcal{H}}]&=&-\frac{i}{2}\sum_{k=1}^{\infty}\frac{(-1)^{k+1}}{k}
\mbox{Tr}\left(G_{\Phi_{0}}\frac{\lambda_{\phi^{2}}}{2}\overline{\mathcal{H}}^{2}\right)^{k}\nonumber\\
&&+\frac{i}{2}\sum_{k=1}^{\infty}\frac{(-1)^{k+1}}{k}\mbox{Tr}(G_{\mathcal{H}'}(3\lambda_{fg}
\overline{\mathcal{H}}^2+\tilde{I}_{2}))^{k}+\tilde{I}_{4}.\nonumber \\
\end{eqnarray}

\begin{figure}
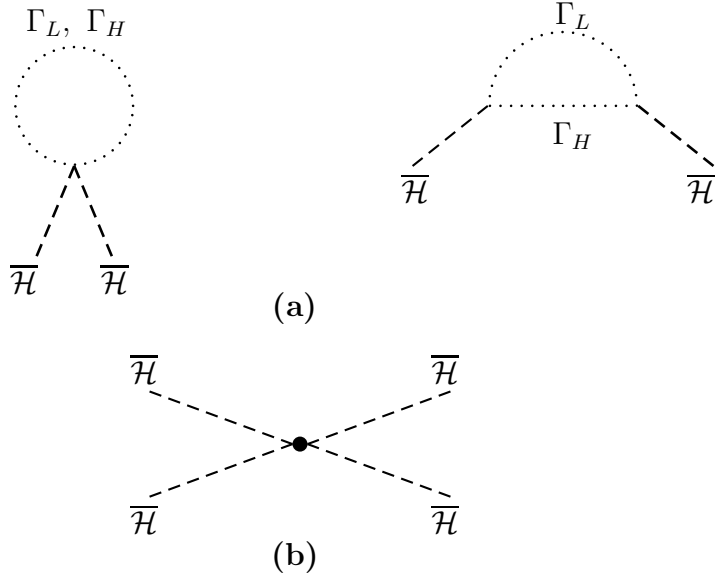

\psset{unit=1mm} 
\psset{arrowsize=2.5} 
\rput (41,1) {$\Gamma_{L},~\Gamma_{H}$}
\rput(40,-10){
\pscircle[linestyle=dotted,linewidth=1pt](0,0){8}
\psline[linestyle=dashed,linewidth=1pt](0,-8)(-5,-20)
\psline[linestyle=dashed,linewidth=1pt](0,-8)(5,-20)
} %
\rput(34,-33){$\mathcal{\overline{H}}$}%
\rput(46,-33){$\mathcal{\overline{H}}$}%
\rput(107,2){$\Gamma_{L}$}
\rput(107,-14){$\Gamma_{H}$}
\rput(105,-10){
\pswedge[linewidth=1pt,linestyle=dotted]{10}{0}{180}
\psline[linestyle=dashed](-10,0)(-20,-8)
\psline[linestyle=dashed,linewidth=1pt](10,0)(20,-8)
}
\rput(86,-21){$\mathcal{\overline{H}}$}
\rput(124,-21){$\mathcal{\overline{H}}$}%
\rput(70,-37){{\bf{(a)}}}%
\rput(50,-45){$\mathcal{\overline{H}}$}
\rput(90,-45){$\mathcal{\overline{H}}$}
\rput(70,-55){
\psline[linestyle=dashed](-20,7)(-1,0)
\psline[linestyle=dashed](-20,-7)(-1,0)
\qdisk(0,0){1}
\psline[linestyle=dashed](20,7)(1,0)
\psline[linestyle=dashed](20,-7)(1,0)
}
\rput(50,-65){$\mathcal{\overline{H}}$}
\rput(90,-65){$\mathcal{\overline{H}}$}
\rput(70,-70){{\bf{(b)}}}%
\vspace{8cm}
\caption{{\bf{(a)}} Scalar sector loops contributing to the Higgs
mass. $\Gamma_{L}=\mathcal{H}', {G^{0}}'$ or $G^{\pm'}$ and
$\Gamma_{H}=\Phi_{0}',\Phi^{P}$   or  $\Phi^{\pm'}$. {\bf{(b)}}
Contribution to the Higgs quartic coupling from the $\Phi_0'$
propagator.}
\label{figuno}
\end{figure}

In order to obtain the scalar contribution to the Higgs mass we only
need to consider the $k=1$ term in the expansion (\ref{Sss}). The
generic loop diagrams are shown in Fig.~\ref{figuno}. Then, for $k=1$,
\begin{eqnarray} \label{muss}
S^{(1)ss}[\mathcal{\overline{H}}]&=&-\frac{i}{2}\lambda_{\phi^{2}}\int dxdy(G_{\Phi_{0}xy}\overline{\mathcal{H}}^{2}\delta_{yx})+\frac{i}{2}\int dxdyG_{\mathcal{H}'xy}(3\lambda_{fg}\overline{\mathcal{H}}^{2}\delta_{yx}+\tilde{I}_{2}\delta_{yx})\nonumber \\
&=&-\frac{i}{4}\lambda_{\phi^{2}}\int dx \overline{\mathcal{H}}^{2}I_{0}(m_{\Phi_{0}}^{2})+\frac{3}{2}i\lambda_{fg}\int dx \overline{\mathcal{H}}^{2}I_{0}(m_{fg}^{2})\nonumber\\
&&+i\lambda_{H^{2}\phi}^{2}f^{2}\int dx \overline{\mathcal{H}}^{2}I_{3}(m_{\Phi_{0}}^{2},m_{fg}^{2})\,,
\end{eqnarray}
with
\begin{eqnarray}
I_{0}(M^{2}) &\equiv& \int d\tilde{k}\frac{i}{(k^{2}-M^{2})}=\frac{1}{(4\pi)^{2}}\left[\Lambda^{2}-M^{2}\log\left(1+\frac{\Lambda^{2}}{M^{2}}\right)\right], \\
I_{3}(M_{a}^{2},M_{b}^{2}) &\equiv& \int d\tilde{p}
\frac{i}{(p^{2}-M_{a}^{2})(p^{2}-M_{b}^{2})}\nonumber\\
&=&-\frac{1}{(4\pi)^{2}}\frac{1}{M_{a}^{2}-M_{b}^{2}}
\left[M_{a}^{2}\log
\left(1+\frac{\Lambda^{2}}{M_{a}^{2}}\right)
- M_{b}^{2}\log\left(1+\frac{\Lambda^{2}}{M_{b}^{2}}\right)\right].\nonumber\\
\end{eqnarray}

For the quartic coupling Higgs correction coming from $\tilde{I}_{4}$ we have:
\begin{equation} \label{lss}
\tilde{I}_{4}=\frac{\lambda_{H^{2}\phi}^{2}}{4\lambda_{\phi^{2}}}\int
dx \overline{\mathcal{H}}^4+...
\end{equation}
where we have expanded the $\Phi_0'$ propagator in powers of
$k^2/m_{\Phi_0'}^2$ and kept just the first term.

\subsection{Pseudoscalar sector and charged sector contributions}

The computation of the contributions from the pseudoscalar and
charged sectors is similar to the previous ones with only one
difference, i.e.: these sectors do not give a contribution to the Higgs
quartic coupling. They just contribute to the Higgs mass. Then the
results for the pseudoscalar sector are:
\begin{eqnarray}\label{mups}
S^{(1)ps}[\mathcal{\overline{H}}]&=&-\frac{i}{4}\lambda_{\phi^{2}}\int dx \overline{\mathcal{H}}^{2}I_{0}(m_{\Phi^{p}}^{2})+\frac{1}{2}i\lambda_{fg}\int dx \overline{\mathcal{H}}^{2}I_{0}(0)\nonumber\\
&+&i\lambda_{H^{2}\phi}^{2}f^{2}\int dx \overline{\mathcal{H}}^{2}I_{3}(m_{\Phi^{p}}^{2},0).
\end{eqnarray}

and for the charged sector:
\begin{eqnarray}\label{mucs}
S^{(1)cs}[\mathcal{\overline{H}}]&=&-\frac{i}{4}\lambda_{\phi^{2}}\int dx \overline{\mathcal{H}}^{2}I_{0}(m_{\Phi^{+}}^{2})+i\lambda_{fg}\int dx \overline{\mathcal{H}}^{2}I_{0}(0)\nonumber\\
&+&i\lambda_{H^{2}\phi}^{2}f^{2}\int dx
\overline{\mathcal{H}}^{2}I_{3}(m_{\Phi^{+}}^{2},0).
\end{eqnarray}
Notice the there is no contribution coming from the doubly
charged scalar sector.

\subsection{Analytical results}

Now by adding (\ref{muss}), (\ref{mups}), (\ref{mucs}) we obtain
the total radiative corrections  to the Higgs mass from the GB
sector up
 to order $\mathcal{O}(v^{2}/f^{2})$ which reads:
\begin{eqnarray} \label{Dm}
\Delta m_{GB}^{2}&=& \frac{3}{(4\pi)^2}\left\{
\left(-\frac{\lambda_{\phi^{2}}}{4}+\lambda_{fg}\right)\Lambda^{2}
+\left(\frac{\lambda_{\phi^{2}}}{4}+\frac{\lambda_{H^{2}\phi}^{2}}{\lambda_{\phi^{2}}}\right)M_{\phi}^{2}\log\left(1+\frac{\Lambda^{2}}{M_{\phi}^{2}}\right)
\right.\nonumber\\
&&\left.-\frac{1}{2}\lambda_{fg}m_{fg}^{2}\log\left(1+\frac{\Lambda^{2}}{m_{fg}^{2}}\right)\right\}\,,
\end{eqnarray}
where, in order to simplify the computations,  we have considered
the heavy scalar fields as
 degenerate  since  $m^{2}/M_{\phi}^{2}$ is of the order of $\mathcal{O}(v^2/f^2)$ (see eq.~(\ref{cs})).

The coefficients of the Higgs potential $\lambda_{fg},
\lambda_{\phi^{2}}$ and  $\lambda_{H^{2}\phi}^{2}$ appearing in
eq.~(\ref{Dm}) receive contributions from both the radiative
corrections and the effective operators (see Appendix). Since the
contributions to
 $\lambda_{fg}$ and $\lambda_{\phi^{2}}$ contain terms of the
order of $\Lambda^{2}$, divergencies $\mathcal{O}(\Lambda^{4})$
and $\mathcal{O}(\Lambda^{2})$ emerge from the first term
in~(\ref{Dm}). However, these divergencies cancel due to the
relationship between $\lambda_{\phi^{2}}$ and $\lambda_{fg}$,
namely:
\begin{eqnarray}\label{rel}
\lambda_{fg}^{\Lambda^{2}} &=& \frac{1}{4} \lambda_{\phi^{2}}^{\Lambda^{2}}, \nonumber\\
\lambda_{fg}^{EO} &=& \frac{1}{4} \lambda_{\phi^{2}}^{EO},
\end{eqnarray}
 where the  index $\Lambda^{2}$ refers to the quadratically
divergent terms and $EO$ represents the part of these coefficients
coming from the effective operators. This fact occurs in the
fermionic and gauge boson sectors, where the quadratic divergences
coming from light and heavy modes of the same statistics
cancel~\cite{Cohen}. Then the corrections summarized in $\Delta
m_{GB}^{2}$ (eq.~\ref{Dm}) are at most of the order
$\mathcal{O}(\Lambda^{2}\log(\Lambda^{2}/M^{2}))$. It is important
to stress that the above cancellations occur exactly only in
\emph{Model I} (as you can easily check from the results given in
the Appendix). However,  in \emph{Model II} (where only the
$SU(2)\times SU(2)\times U(1)$ is gauged), there are
$\mathcal{O}(\Lambda^{2})$ terms coming from the U(1) sector which
do not cancel. However, such terms appear always with  a squared
gauge coupling $g'$ factor which is very small ($g'^2/g^2 \sim
0.3$ in the SM) and then their contribution is not expected to be
too large.

Finally, from (\ref{lss}), the radiative correction to the quartic
coupling  is:
\begin{eqnarray} \label{lGB}
\tilde{I}_{4}=\frac{1}{4}
\Delta \lambda_{GB} \int dx \overline{\mathcal{H}}^4\,,\nonumber
\end{eqnarray}
being
\begin{eqnarray}
\Delta \lambda_{GB}=
\frac{\lambda_{H^{2}\phi}^{2}}{\lambda_{\phi^{2}}}.
\end{eqnarray}

In summary, taking into account (\ref{lss}) and (\ref{Dm}), the
Higgs boson potential can be written as:
\begin{equation}
V=\frac{1}{2}m_{\overline{\mathcal{H}}}^{2}{\overline{\mathcal{H}}}^{2}+\frac{1}{4}\lambda_{\overline{\mathcal{H}}}{\overline{\mathcal{H}}}^{4},
\end{equation}
where the Higgs mass is given by,
\begin{equation} \label{ml}
m_{\overline{\mathcal{H}}}^{2}=2(\mu_{fg}^{2}-\Delta m_{GB}^{2}),
\end{equation}
and the quartic Higgs couplings is,
\begin{equation}
\lambda_{\overline{\mathcal{H}}}=
\lambda_{fg}-\frac{\lambda_{H^{2}\phi}^{2}}{\lambda_{\phi^{2}}}.
\end{equation}

It is important to note that we have obtained the GB contributions
after having broken the SM symmetry through the fermion and
gauge boson radiative corrections. In this fact we differ from
other analysis performed in the literature (see for example
\cite{Cohen,Casas}), where these scalar contributions are computed
at the tree level from the effective operators only.
  Moreover, in our case, the coefficients of the potential (\ref{potef1})
do not depend only on the two unknown coefficients $a$ and $a'$, but also
on the scale $f$ and the cutoff $\Lambda$, thus setting more
restrictions on the space parameter as we will see in the following.

\section{Numerical Results and Phenomenological Discussion}

In this section we continue our study about the allowed region of
the parameter space of the LH model started in our previous
papers~\cite{ATP,ATP2,ATP3}. In the present one we complete this
phenomenological study, taking into account also the contributions
from the Goldstone boson sector to the Higgs mass and  quartic
coupling obtained above. The LH parameters different relationships
and their relevant ranges considered are the following:

First, we impose the minimum condition for the complete effective
potential (\ref{potef1}):
\begin{equation}
\label{eq:mincond}
v^2=\frac{\mu_{fg}^{2}}{\lambda_{fg}-\lambda_{H^{2}\phi}^{2}/\lambda_{\phi^{2}}}\,.
\end{equation}
This condition is crucial in order to reproduce the electroweak
symmetry breaking.

If we want to study the allowed region of the parameter space in these models,
we should also take into account other constraints imposed by requiring the
consistency of the LH models with the electroweak precision data.
 There exist several studies of the corrections to
electroweak precision observables in the Little Higgs models,
exploring whether there are regions of the parameter space in
which the model is consistent with the available
data~\cite{Logan,Peskin,Csaki1,Csaki2,Schmaltz,review1,LoganP,EWPO1,recentpheno}.
In \emph{Model I} with a gauge group $SU(2)\times SU(2) \times
U(1) \times U(1)$ we have a multiplet of heavy $SU(2)$ gauge
bosons and a heavy $U(1)$ gauge boson. The last one leads to large
electroweak corrections and some problems with the direct
observational bounds on the $Z'$ boson from
Tevatron~\cite{Csaki1,Csaki2}. Then, a very strong bound on the
symmetry breaking scale $f$, $f>4$ TeV at $95 \%$ C.L, is
found~\cite{Csaki1}. However, it is known that this bound is
lowered to $1-2$ TeV for some region of the parameter
space~\cite{Csaki2} by gauging only $SU(2)\times SU(2) \times
U(1)$ (\emph{Model II}). For this reason, in the following we will
concentrate only on this model.

On the other hand, in order to avoid small values for the $W'$
mass and a very strong coupling constant, we set the range of the
$\psi$ mixing  angle (for the $SU(2)$ group) to be
$0.1<c{_\psi}<0.9$~\cite{ATP2}. In addition, the condition
$\lambda_T \gsim$ 0.5 is established  from the top mass
~\cite{Logan}, setting the bounds on the couplings $\lambda_1,
\lambda_2 \geq m_t/v$ or $\lambda_1 \lambda_2 \geq 2 (m_t/v)^2$.
In order to avoid a large fine-tuning in the Higgs potential
~\cite{Cohen,Peskin} we set the condition  $m_T\lsim$ 2.5 TeV.
Then, since $m_{T}$ grows linearly with $f$, $f$ should be less
than about one TeV~\cite{ATP}. Following the restrictions on the
parameters given in~\cite{ATP2}, we take  $0.8$ TeV $<f<1$ TeV.
Finally the usual condition $\Lambda \lsim 4 \pi f$ is also
imposed.

By using the constraints on  the LH parameters given above, taking
into account also that the Higgs mass is experimentally restricted
to the range $114$ GeV $< m_{\overline{\mathcal{H}}}<200$ GeV, and
by imposing the minimum condition (\ref{eq:mincond}), we analyze
the available regions for the remaining LH parameters. To do that
we include the contributions of both radiative corrections and
effective operators. In fact, in order to see the role played for
each of them, we consider three different cases: having just
radiative corrections (RC), just effective operators (EO) and the
most general case including both of them (RC+EO).

\begin{figure}[h!]
\rput(3,-2.3){\epsfig{file=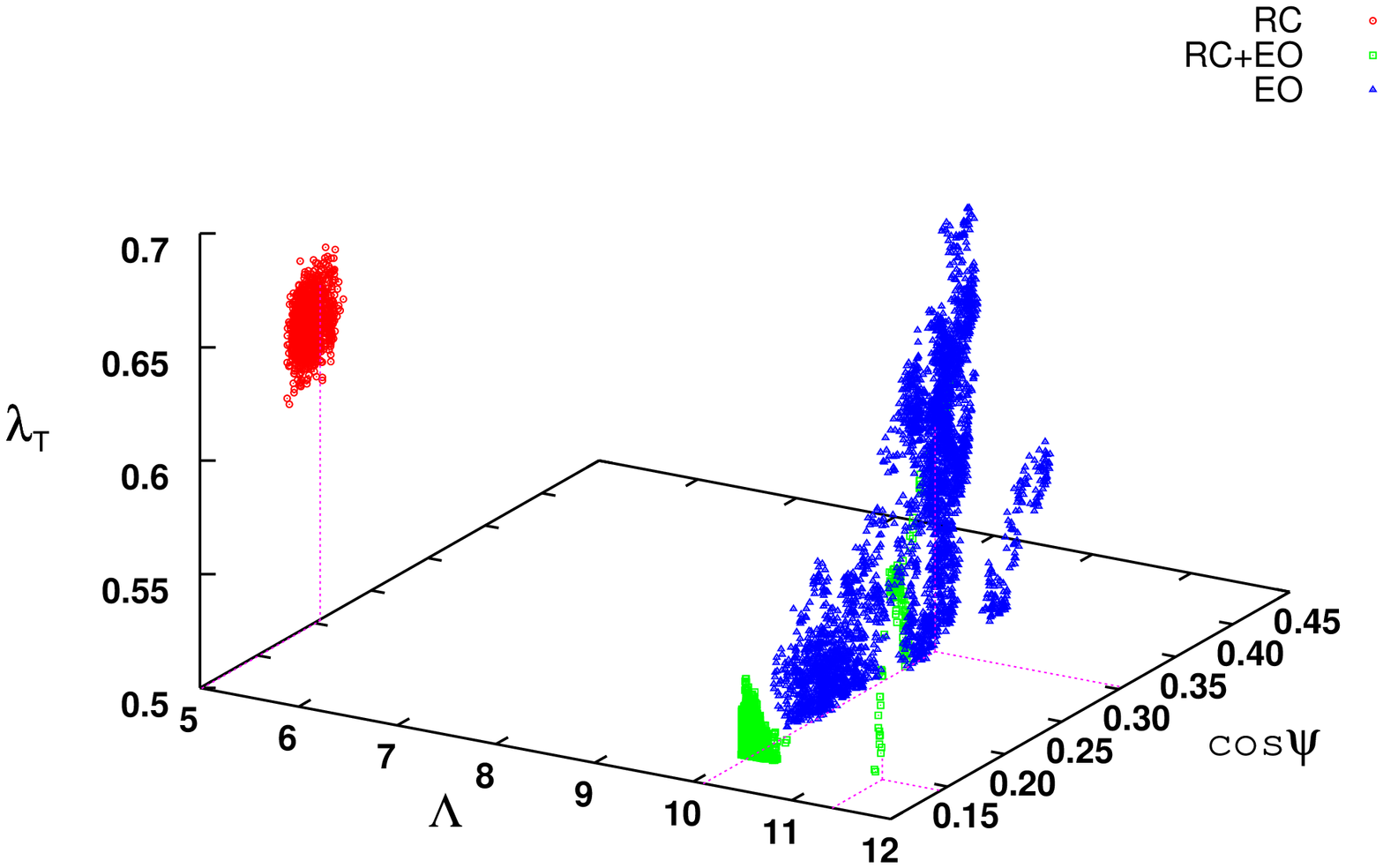,scale=0.35,angle=270}}%
\rput(10,-2.5){\epsfig{file=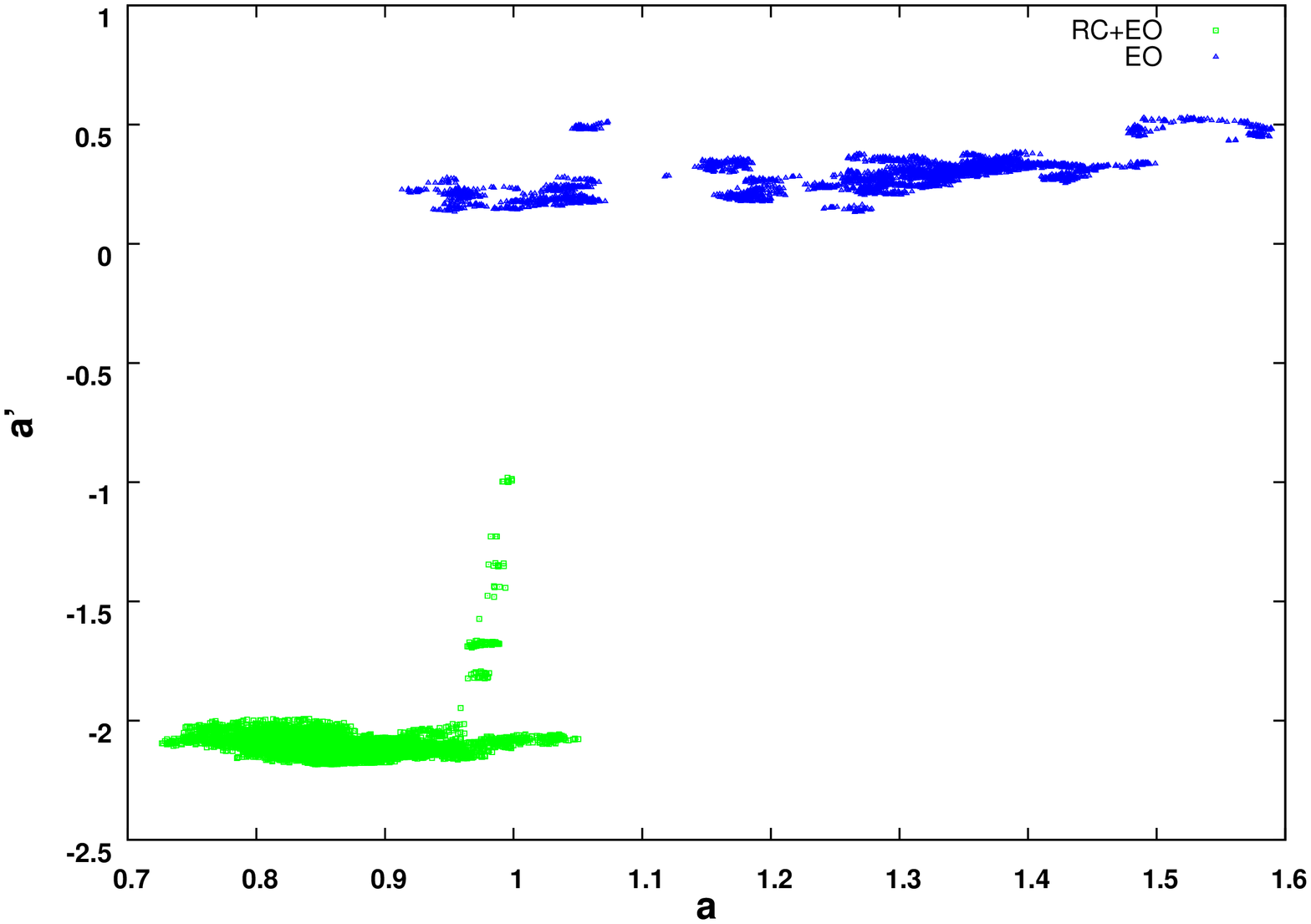,scale=0.3,angle=270}}%
\vspace*{5cm}
\caption{{\bf{(a)}} Values of $\lambda_T$, $\Lambda$ and $c_{\psi}$ which are
 possible solutions for the LH model. Here $f$ vary between $0.8$ and $1$ TeV, and
  $a$ and $a'$ are $\mathcal{O}(1)$. The three separate surfaces correspond with
the three different cases analyzed in this section.
 {\bf{(b)}} Values of $a$ and $a'$ which are possible solutions for the LH model.
 The $\lambda_T$, $\Lambda$, $c_{\psi}$ and $f$ ranges are described in the text.}
\label{surface}
\end{figure}

\begin{figure}[h!]
\rput(2.8,-2.5){\epsfig{file=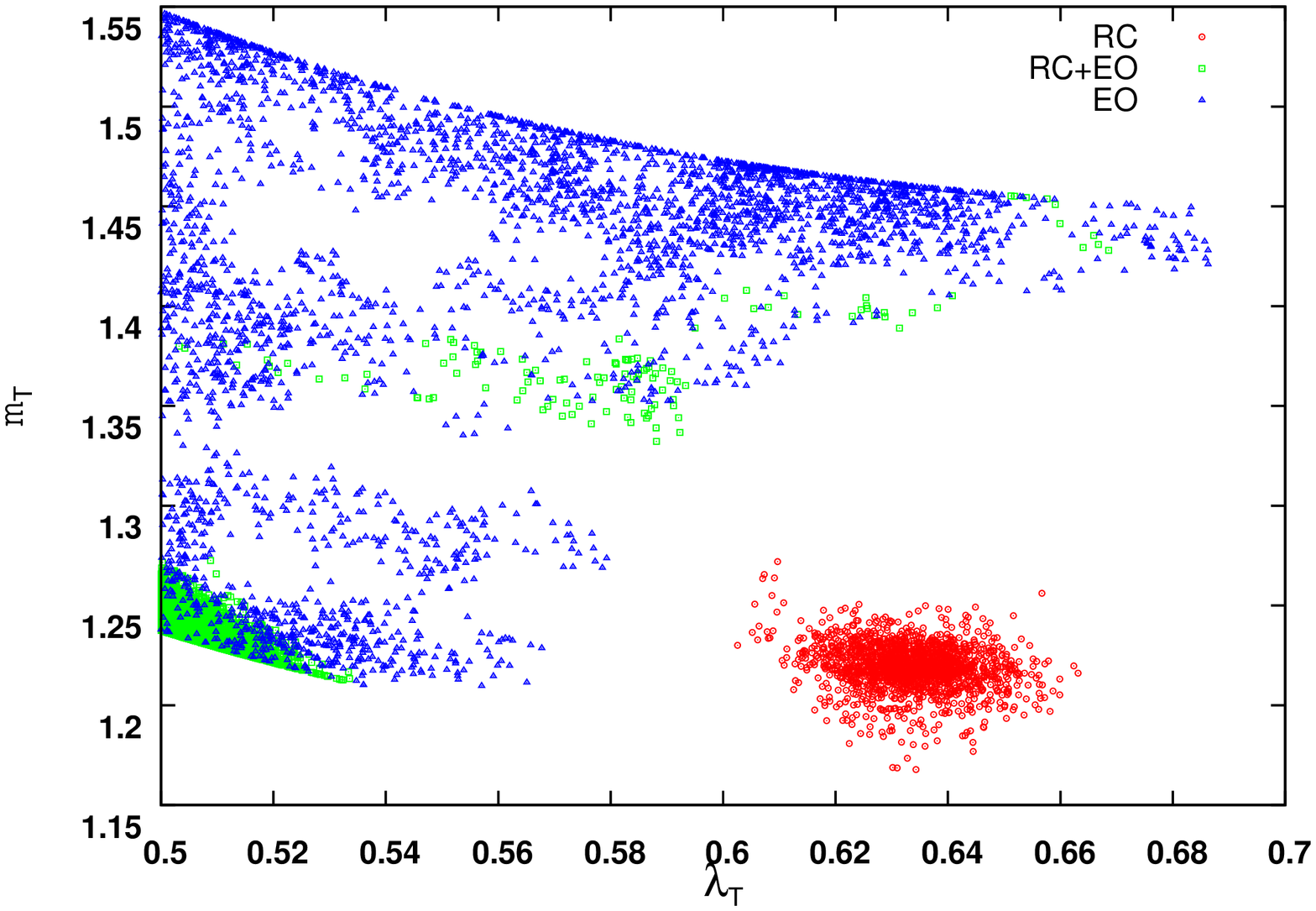,scale=0.3,angle=270}}%
\rput(10,-2.5){\epsfig{file=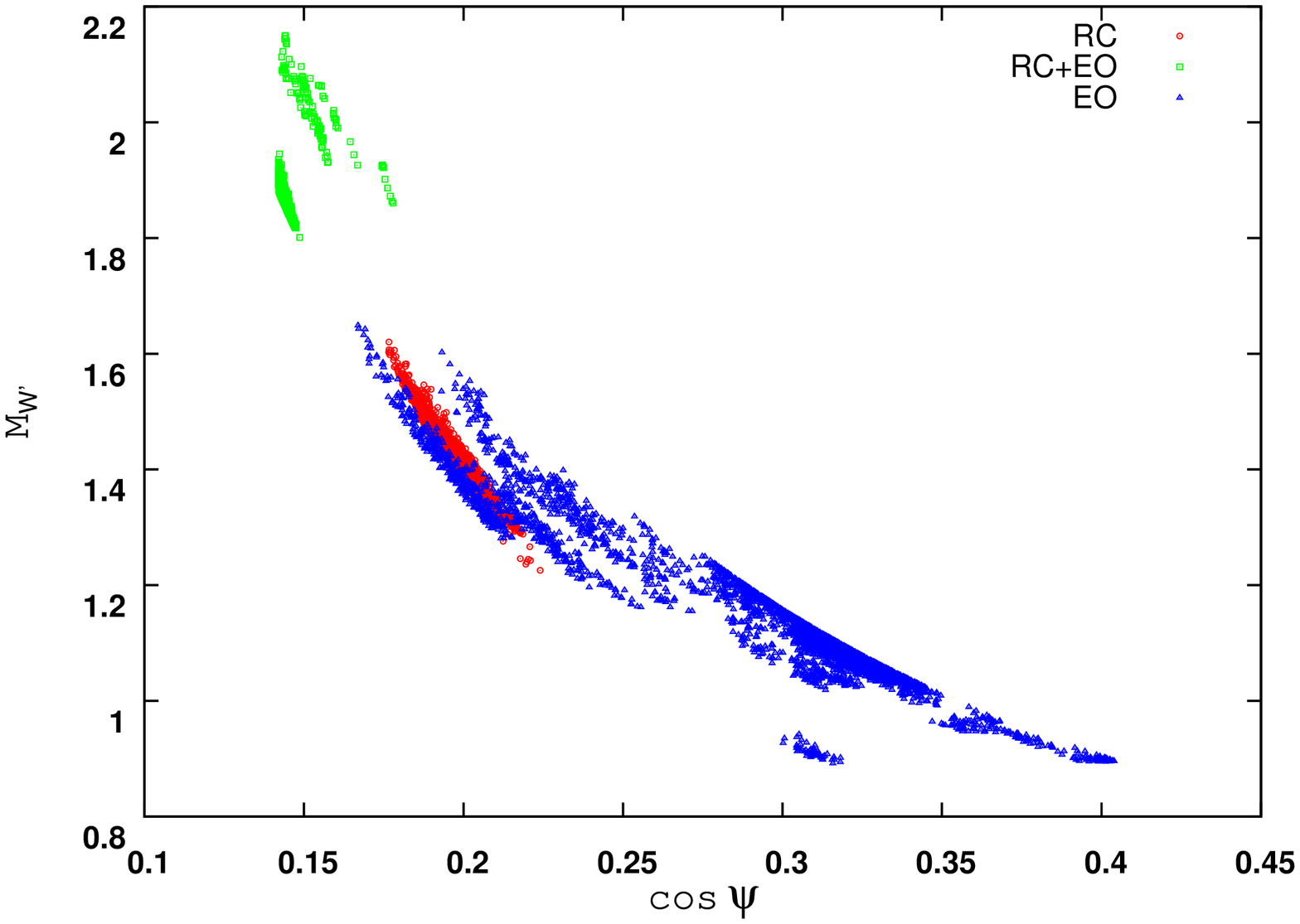,scale=0.3,angle=270}}%
\rput(7,-8){\epsfig{file=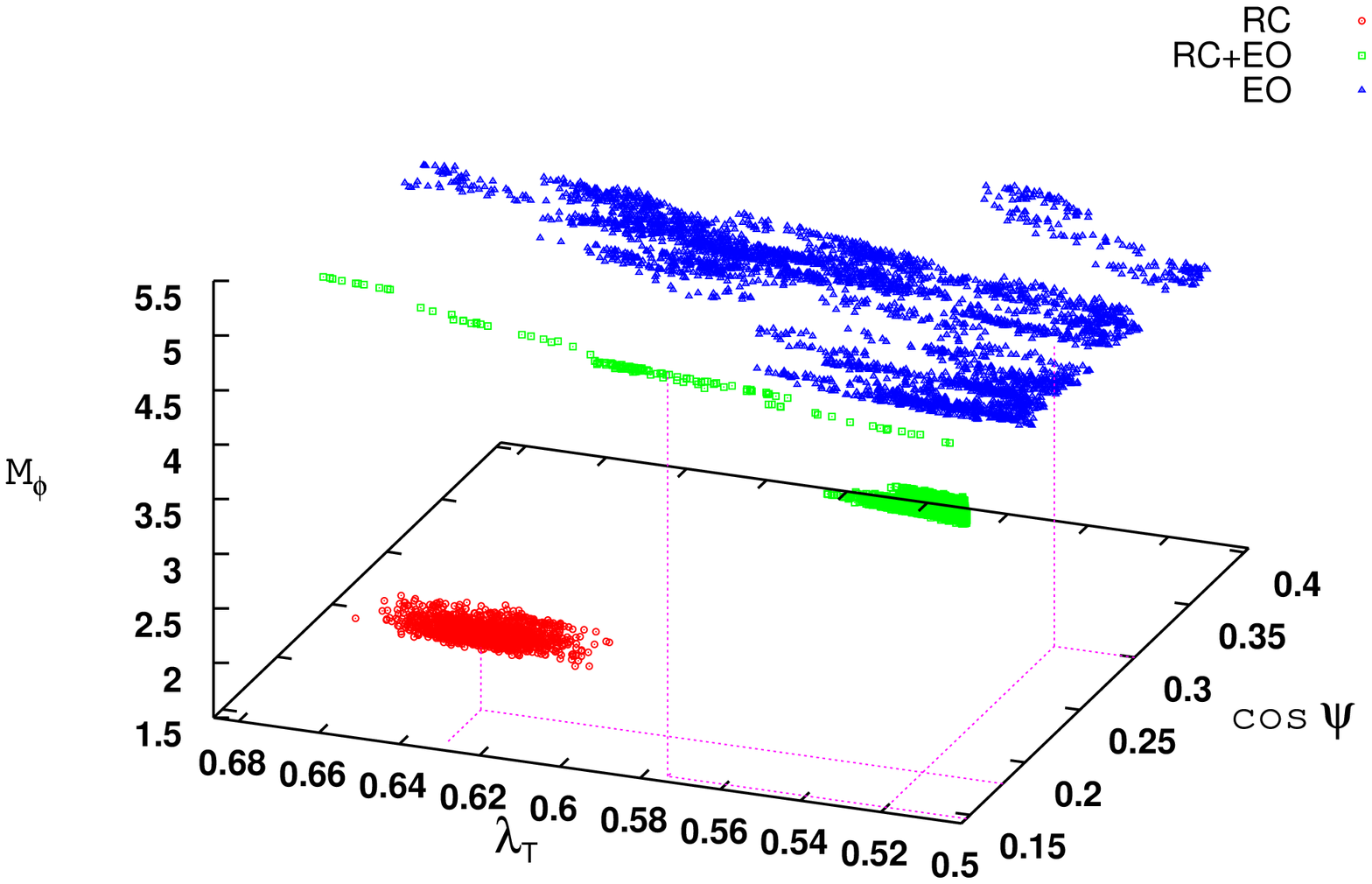,scale=0.37,angle=270}}%
\vspace*{11cm}
\caption{{\bf{(a)}} $m_{T}$ as a function of $\lambda_{T}$,
 {\bf{(b)}} $M_{W'}$ as a function of $\cos \psi$ and {\bf{(c)}}
 $M_{\phi}$ as a function of $\lambda_T$ and $\cos \psi$, where the $\Lambda$,
  $f$, $a$ and $a'$ parameters vary between ranges described in the text.\vspace{-2ex}}
\label{masses}
\end{figure}

In Fig.~\ref{surface} we show the allowed regions of the parameter
space for the three different cases analyzed;  RC (red region), EO
(blue region) and RC+EO (green region). In Fig.~2.a we show the
possible solutions to the LH model in the $(\Lambda, c_{\psi},
\lambda_{T})$ space varying $f$ between $0.8$ TeV and $1$ TeV and
by assuming that the $a$ and $a'$
 parameters are of the order of $\mathcal{O}(1)$.
From these results there are two important issues to remark.
First, when only radiative corrections are included we do not find
any solution for the LH model if $\Lambda > 6$ TeV. Unfortunately,
precision electroweak data rule out new strong interactions at
scales below about $10$ TeV. On the contrary, in the other two
cases, RC+EO and EO, the possible values for the cut-off are
larger. This fact implies also that the mass of the $\phi$
 fields must be about $2$ TeV when the model includes only radiative corrections unlike in
  the other two cases where it is about $5$ TeV
(see Fig.~\ref{masses}). In Fig.~2.b we show
 the possible values for the unknown $a$ and $a'$ parameters.
Here, the other parameters have been varied in the ranges set
above. The two cases  considered are RC+EO and EO only. We find
that the set of possible solutions include in both cases positive
 values for $a$.   In the RC+EO case we obtain
 large and negative  values for  $a'$, whereas
  in the EO case $a'$ takes small and positive values. Notice also that $a$ is always
   positive. This is important since it is known that $a<0$ leads to
a large {\it{v.e.v}} for the scalar triplet.

\begin{figure}[h!]
\rput(2.7,-2.5){\epsfig{file=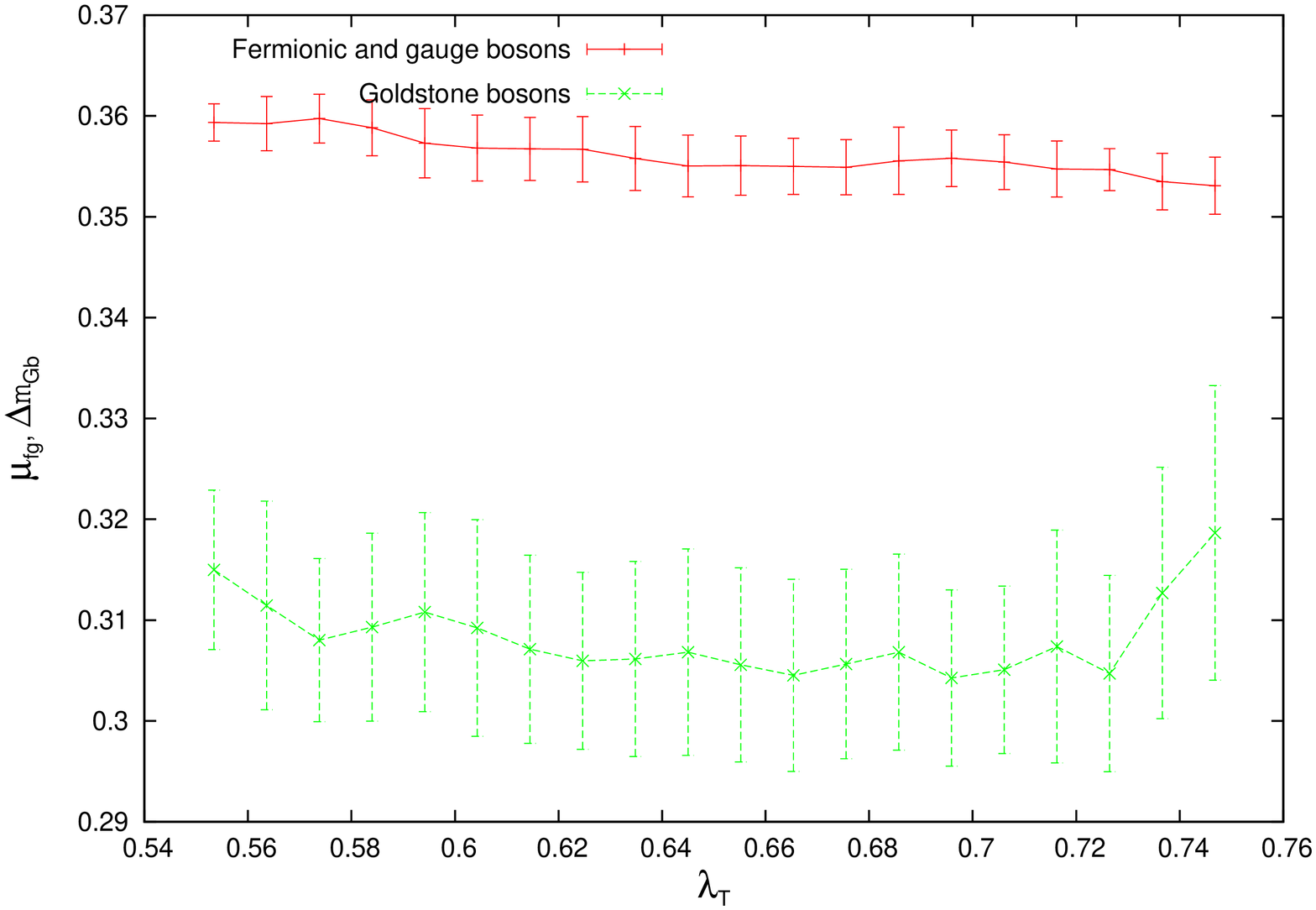,scale=0.3,angle=270}}%
\rput(10.5,-2.5){\epsfig{file=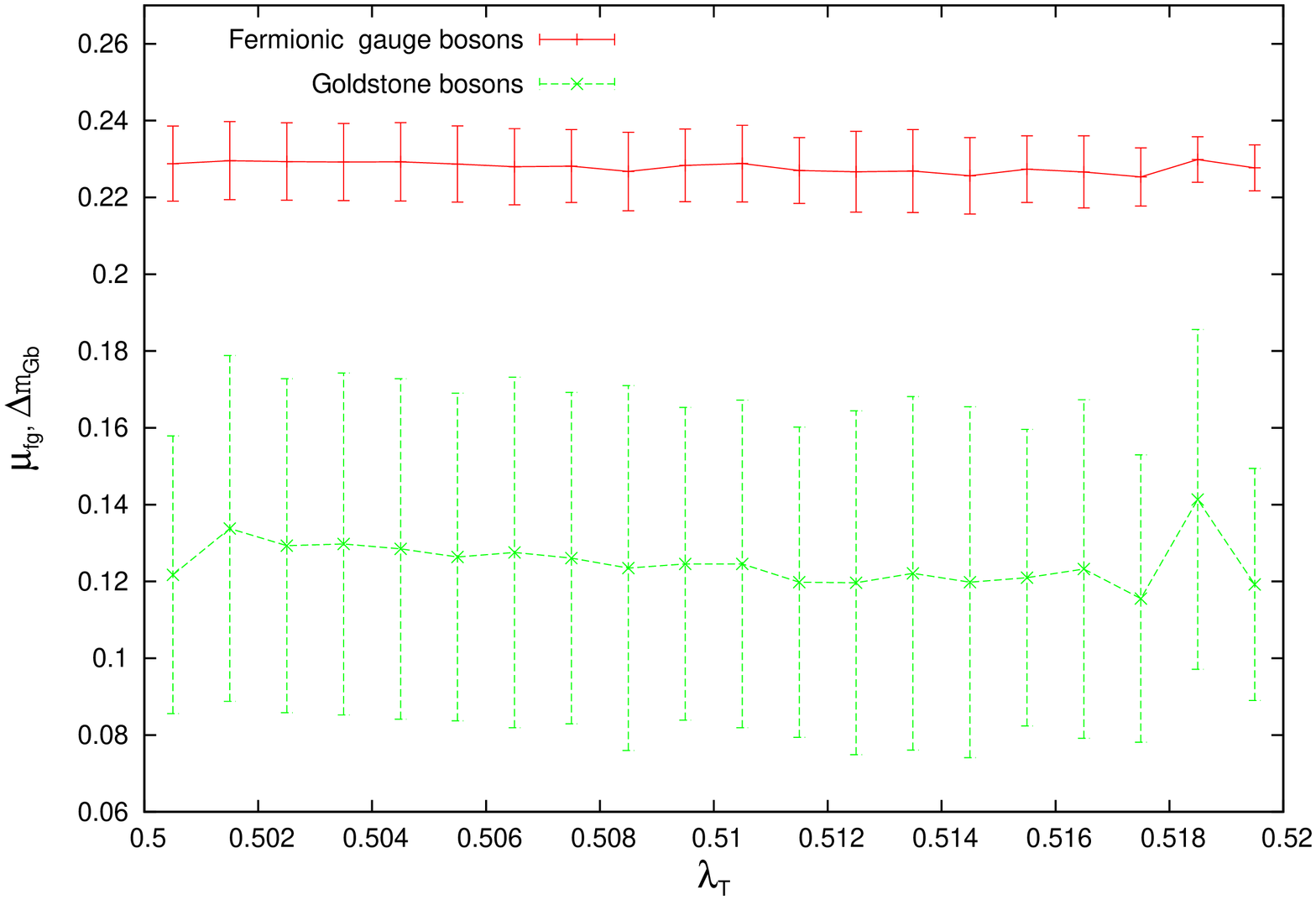,scale=0.3,angle=270}}%
\rput(7,-8.5){\epsfig{file=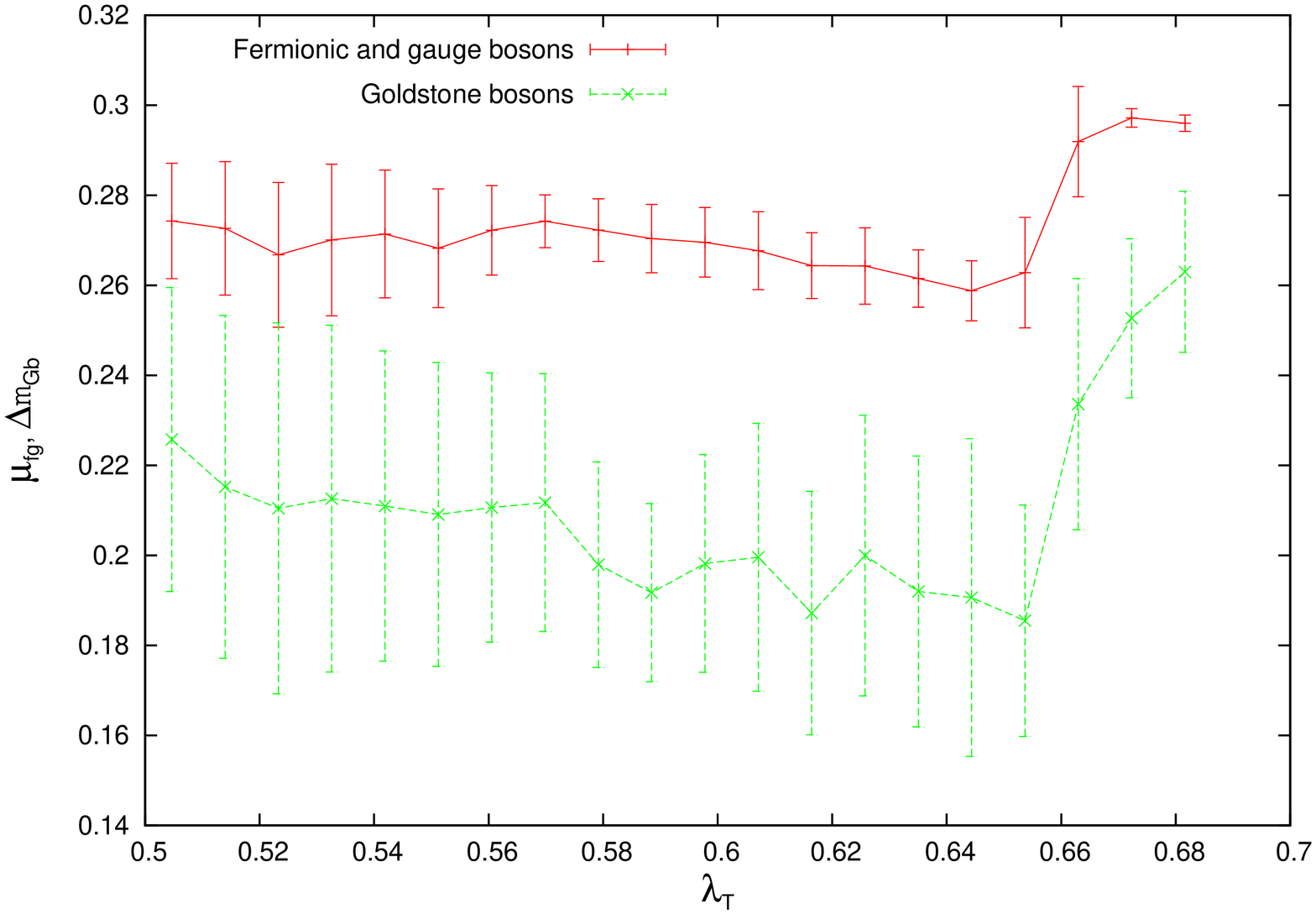,scale=0.3,angle=270}}%
\vspace*{11cm} \caption{These figures show the average and
standard deviation of both the fermionic and gauge boson
contribution $\mu_{fg}$ and the Goldstone boson contribution
$\Delta m_{GB}$ to the Higgs mass as a function of $\lambda_T$.
{\bf{(a)}} RC case,
 {\bf{(b)}} RC+EO case and {\bf{(c)}} EO case.}
\label{compare}
\end{figure}

The reason for the differences of the parameter solutions for the
three cases come from the
  $\Delta m_{GB}^{2}$    cutoff dependence when the radiative contributions
  are included. For example, in the case where only the radiative
corrections are  taken into account, a cut-off $\Lambda$  bigger
than $6$ TeV produces GB contributions  resulting in a  negative
Higgs mass. However, by  dropping
  the value of $\Lambda$ we get a LH parameter space where
the condition~(\ref{eq:mincond}) is satisfied and the Higgs
       mass is well inside the experimental constraints.
       In the RC+EO case, the $a'$ parameter can take values
        which help to compensate the big effect of the GB radiative contributions
(see also Fig.~\ref{compare}) thus allowing larger cutoff values.

For completeness, Fig.~\ref{masses} shows the mass values for the
heavy particles in the
 three different cases analyzed. Each point of the figures is a possible solution
 of the LH model. In this way, these regions represent the possible values for the masses of
  the heavy particles predicted by the LH model, which are  compatible with  electroweak symmetry breaking
   and precision data.
The region of possible values for the masses coming from EO
contributions is clearly larger than in the case of considering
RC alone. Notice that the theoretical lower bounds in the heavy
states masses, $M_\phi, M_{W'} \gsim $ 1 TeV, and the condition
$m_T\lsim$ 2.5 TeV are fulfilled.

To complete our study, we compare the contributions to the Higgs
mass coming from the different sectors i.e. fermionic and gauge
bosons ($\mu_{fg}$) and on the other hand the GB contribution
($\Delta m_{GB}$), as a function of $\lambda_T$. We show the
average and standard deviation for each contribution
(Fig.~\ref{compare}). In all physical cases it can be seen that
$\mu_{fg}>\Delta m_{GB}$, thus yielding a real value for the Higgs
mass (eq.~\ref{ml}). It is also remarkable
 the higher variability of $\Delta m_{GB}$ compared with
 $\mu_{fg}$. The reason is that both the parameters appearing
  in the radiative corrections, i.e. $f, \Lambda, \lambda_T, \cos \phi$, and the two EO
  parameters
$a$ and $a'$, play an important role in the final results of $\Delta m_{GB}$
(see the discussion above).

Finally, as an example, we give in the  Table.1 the lowest Higgs
mass values found for the three cases considered in this work.

\begin{center}
\begin{table}[h!]
\begin{tabular}{||l|l|l|l||}
\hline
\hline
\textbf{Parameters} & \textbf{RC} & \textbf{RC+EO} & \textbf{EO} \\
\hline
$m_{\overline{\mathcal{H}}}$ & $156.66$ GeV & $114.69$ GeV & $116.94$ GeV\\
\hline
$\mu_{fg}$ & $359.54$ GeV & $236.87$ GeV & $288.70$ GeV\\
\hline
$\Delta_{Gb}$ & $342.04$ GeV & $222.55$ GeV & $275.53$ GeV\\
\hline
$\lambda_{\overline{\mathcal{H}}}$ & $0.97$ & $0.90$ & $1.42$\\
\hline
$f$ & $0.86$ TeV & $0.96$ TeV & $0.82$ TeV\\
\hline
$\Lambda$ & $5$ TeV & $11.64$ TeV & $10.01$ TeV\\
\hline
$\lambda_{T}$ & $0.6$ & $0.61$ & $0.53$\\
\hline
$c_{\psi}$ & $0.18$ & $0.16$ & $0.3$\\
\hline
$a$ & $0$ & $0.98$ & $1.06$\\
\hline
$a'$ & $0$ & $-1.25$ & $0.5$\\
\hline
\hline
\end{tabular}\label{Table}
\caption{The lowest values for the Higgs mass found for the three cases: RC, RC+EO and EO.}
\end{table}
\end{center}


\section{Conclusion}

In this work we have completed our program of computing the
relevant contributions to the Higgs low-energy effective potential
in the context of the Littlest Higgs models based on the
$SU(5)/SO(5)$ coset. To the radiative corrections coming from the
fermions and the gauge bosons considered so far, we have added
here the effect the scalar loops and also the effective operators
emerging from the ultraviolet completion of the model.

In particular we have computed in detail the main contributions to
the Higgs mass and its quartic coupling. From our previous works,
in which only fermionic and gauge boson radiative corrections were
included, it was clear that the effect of the scalar sector could
be decisive in order to have the appropriate cancellations between
the different sectors of the model to give a Higgs mass within the
present experimental limits. We have performed  our analytical
computations for two different versions of the model called
\emph{Model I} and \emph{Model II} having as gauge groups
$[SU(2)\times U(1)]^2$ and $SU(2)^2\times U(1)$ respectively.

In order to complete our analysis, we have concentrated on
studying those regions of the parameter space where the model
could give rise to an acceptable phenomenology. In particular we
have done a detailed numerical search for \emph{Model II} since
\emph{Model I} seems to be  incompatible with the present
experimental data \cite{Csaki1,Csaki2}. We have analyzed three
cases: 1)~radiative corrections only (RC), 2)~radiative corrections
and effective operators
  (RC+EO) and 3)~effective operator only (EO). From this analysis
  we get that this  model is compatible with the expected
  Higgs mass provided that the contribution of the effective operators is
  included. We also conclude that the Goldstone boson contributions are
fundamental to obtain a low enough Higgs particle mass. For
example  a Higgs mass $m_H \simeq 115 GeV$ can be obtained when
radiative and effective operator contributions are both taken into
account.

Summarizing, we have arrived to the conclusion that the $SU(5)/S(5)$
Littlest Higgs model with gauge group $[SU(2)\times U(1)]^2$ is
phenomenologically viable through some tuning in the parameter space,
assuming a careful inclusion of fermions, gauge bosons, scalar
loops and effective operators.

In any case it will be the LHC, whose main goal is to disentangle
the mechanism of the electroweak symmetry breaking, which will
decide if Littlest Higgs models are appropriate for describing
mechanism or not.

\vspace{1cm} {\bf Acknowledgments:} This work is supported by
DGICYT (Spain) under project number FPA2008-00592 and by the
Universidad Complutense/CAM: UCM-BSCH GR58/08 910309. The work of
S.P.\ is supported by a {\it Ram{\'o}n y Cajal} contract from MEC
(Spain) and partially by CICYT (grant FPA2006-2315) and DGIID-DGA
(grant 2008-E24/2).The work of J.R.L. is supported by project
number FIS2006-04885. We would like to thank J.R.Espinosa for
useful discussions.

\newpage
\textbf{\appendixname } \vspace{0.3cm}

\textbf{a. Coefficients coming from loops computation}

\vspace{0.3cm}

\emph{Model I}

\begin{eqnarray}
\mu_{fg}^{2} &=& \mu^{2}_f+\mu^{2}_g  \nonumber \\
&=& N_{c} \frac{m_{T}^{2} \lambda_{t}^{2}}{4 \pi^{2}}
\log\left(1+\frac{\Lambda^{2}}{m_{T}^{2}}\right)\nonumber\\
&-&\frac{3}{64\pi^{2}}\left[3g^{2}M_{W'}^{2}
\log\left(1+\frac{\Lambda^{2}}{M_{W'}^{2}}\right)+g^{'2}M_{B'}^{2}\log\left(1+\frac{\Lambda^{2}}{M_{B'}^{2}}
\right)\right] \nonumber
\end{eqnarray}

\begin{eqnarray}
\lambda_f&=&\frac{N_{c}}{(4
\pi)^{2}}\left[2(\lambda_{t}^{2}+\lambda_{T}^{2})
\frac{\Lambda^{2}}{f^{2}}\right.\nonumber\\
&-&\log\left(1+\frac{\Lambda^{2}}{m_{T}^{2}}\right)
\left(-\frac{2m_{T}^{2}}{f^{2}}\left(\frac{5}{3}
\lambda_{t}^{2}+\lambda_T^{2}\right)+4\lambda_{t}^{4}
+4(\lambda_{T}^{2}+
\lambda_{t}^{2})^{2}\right)\nonumber\\
&-&\left.4\lambda_{T}^{2}\frac{1}{1+\frac{m_{T}^{2}}{\Lambda^{2}}}\left
(\frac{m_{T}^{2}}{f^{2}}-2\lambda_{t}^{2} -\lambda_{T}^{2}\right)
-4\lambda_{t}^{4}
\log\left(\frac{\Lambda^{2}}{m^{2}}\right)\right]\nonumber \\
&-&\frac{3}{(16\pi f)^{2}}\left[-\left(\frac{g^{2}}{c_{\psi}^{2}s_{\psi}^{2}}+
\frac{g^{'2}}{c_{\psi}^{'2}s_{\psi}^{'2}}\right)\Lambda^{2}\right.\nonumber\\
&+&\left.g^{2}M_{W'}^{2}
\log\left(1+\frac{\Lambda^{2}}{M_{W'}^{2}}\right)\left(4+\frac{1}{c_{\psi}^{2}s_{\psi}^{2}}
+2g^{'2}\frac{(c_{\psi}^{2}s_{\psi}^{'2}+s_{\psi}^{2}c_{\psi}^{'2})^{2}}{c_{\psi}^{2}
s_{\psi}^{2}c_{\psi}^{'2}s_{\psi}^{'2}}\frac{f^{2}}{M_{W'}^{2}-M_{B'}^{2}}\right)
\right.\nonumber \\
&+&\left.g^{'2}M_{B'}^{2} \log\left(1+\frac{\Lambda^{2}}{M_{B'}^{2}}\right)
\left(\frac{4}{3}+\frac{1}{c_{\psi}^{'2}s_{\psi}^{'2}}
+2g^{2}\frac{(c_{\psi}^{2}s_{\psi}^{'2}+s_{\psi}^{2}c_{\psi}^{'2})^{2}}{c_{\psi}^{2}
s_{\psi}^{2}c_{\psi}^{'2}s_{\psi}^{'2}}\frac{f^{2}}{M_{B'}^{2}-M_{W'}^{2}}\right)
\right.\nonumber \\
&+&\left.f^{2}\log\left(1+\frac{\Lambda^{2}}{M_{W'}^{2}}\right)\left(3g^{4}+
2(3g^{2}+g^{'2})g^{2}\frac{(s_{\psi}^{2}-c_{\psi}^{2})^{2}}{c_{\psi}^{2}s_{\psi}^{2}}\right)\right.\nonumber\\
&+&\left.f^{2}\log\left(1+\frac{\Lambda^{2}}{M_{B'}^{2}}\right)\left(g^{'4}+2(g^{2}+
g^{'2})g^{'2}\frac{(s_{\psi}^{'2}-c_{\psi}^{'2})^{2}}{c_{\psi}^{'2}s_{\psi}^{'2}}
\right)\right.\nonumber \\
&+&\left.f^{2}\log\left(\frac{\Lambda^{2}}{m^{2}}\right)\left(3g^{4}+g^{'4}+
8g^{2}g^{'2}\right)-3f^{2}\frac{g^{4}}{1-\frac{M_{W'}^{2}}{\Lambda^{2}}}-
f^{2}\frac{g^{'4}}{1-\frac{M_{B'}^{2}}{\Lambda^{2}}}\right]\nonumber
\end{eqnarray}

\begin{eqnarray}
\lambda_{\phi^2f}&=&\frac{8N_{c}}{(4\pi f)^{2}}(\lambda_{t}^{2}+\lambda_{T}^{2})\left(\Lambda^{2}-m_{T}^{2}
\log\left(\frac{\Lambda^{2}}{m_{T}^{2}}+1\right)\right)\nonumber\\
&+&\frac{3}{4(4\pi f)^2}\left[\frac{g^{2}}{c_{\psi}^{2}s_{\psi}^{2}}\Lambda^{2}-g^{2}M_{W'}^{2}\log\left(\frac{\Lambda^{2}}{M_{W'}^{2}}+1\right)\left(\frac{(s_{\psi}^{2}-c_{\psi}^{2})^{2}}{c_{\psi}^{2}s_{\psi}^{2}}-4 \right)\right.\nonumber\\
&+&\left.\frac{g^{'2}}{c_{\psi'}^{2}s_{\psi'}^{2}}\Lambda^{2}-g^{'2}M_{B'}^{2}\log\left(\frac{\Lambda^{2}}{M_{B'}^{2}}+1\right)\frac{(s_{\psi'}^{2}-c_{\psi'}^{2})^{2}}{c_{\psi'}^{2}s_{\psi'}^{2}}\right]\nonumber
\end{eqnarray}

\begin{eqnarray}
\lambda_{H^{2}\phi}&=&-\frac{4N_{c}}{(4\pi f)^{2}}\left[(\lambda_{t}^{2}+\lambda_{T}^{2})\Lambda^{2}-\lambda_{T}^{2}m_{T}^{2} \log\left(\frac{\Lambda^{2}}{m_{T}^{2}}+1\right)\right]\nonumber \\
&+&\frac{3}{8(4\pi f)^{2}}\left[g^{2}\frac{s_{\psi}^{2}-c_{\psi}^{2}}{c_{\psi}^{2}s_{\psi}^{2}}\left(\Lambda^{2}-M_{W'}^{2}\log\left(\frac{\Lambda^{2}}{M_{W'}^{2}}+1\right)\right)\right.\nonumber\\
&+&\left.g^{'2}\frac{s_{\psi'}^{2}-c_{\psi'}^{2}}{c_{\psi'}^{2}s_{\psi'}^{2}}\left(\Lambda^{2}-M_{B'}^{2}\log\left(\frac{\Lambda^{2}}{M_{B'}^{2}}+1\right)\right)\right]\,,\nonumber
\end{eqnarray}

\vspace{0.5cm}

\emph{Model II}
\begin{eqnarray}
\mu_{fg}^{2} &=& \mu^{2}_f+\mu^{2}_g  \nonumber \\
&=& N_{c} \frac{m_{T}^{2} \lambda_{t}^{2}}{4 \pi^{2}}
\log\left(1+\frac{\Lambda^{2}}{m_{T}^{2}}\right)-\frac{3}{64\pi^{2}}\left(3g^{2}M_{W'}^{2}
\log\left(1+\frac{\Lambda^{2}}{M_{W'}^{2}}\right)+g^{'2} \Lambda^{2}\right)\nonumber
\end{eqnarray}

\begin{eqnarray}
\lambda_{fg}&=&\frac{N_{c}}{(4
\pi)^{2}}\left[2(\lambda_{t}^{2}+\lambda_{T}^{2})
\frac{\Lambda^{2}}{f^{2}}\right.\nonumber\\
&-&\log\left(1+\frac{\Lambda^{2}}{m_{T}^{2}}\right)
\left(-\frac{2m_{T}^{2}}{f^{2}}\left(\frac{5}{3}
\lambda_{t}^{2}+\lambda_T^{2}\right)+4\lambda_{t}^{4}
+4(\lambda_{T}^{2}+
\lambda_{t}^{2})^{2}\right)\nonumber\\
&-&\left.4\lambda_{T}^{2}\frac{1}{1+\frac{m_{T}^{2}}{\Lambda^{2}}}\left
(\frac{m_{T}^{2}}{f^{2}}-2\lambda_{t}^{2} -\lambda_{T}^{2}\right)
-4\lambda_{t}^{4}
\log\left(\frac{\Lambda^{2}}{m^{2}}\right)\right]\nonumber\\
&-&\frac{3}{(16 \pi f)^{2}}
\left[-\frac{g^{2}}{c_{\psi}^{2}s_{\psi}^{2}}\Lambda^{2}+
\frac{4}{3}{g'}^{2}\Lambda^{2}+ g^{2}M_{W'}^{2}
\log\left(\frac{\Lambda^{2}}{M_{W'}^{2}}+1\right)
\left(4+\frac{1}{c_{\psi}^{2}s_{\psi}^{2}}\right) \right. \nonumber \\
&+&\left.f^{2}\log \left(1+\frac{\Lambda^{2}}{M_{W'}^{2}}\right)
\left(3g^{4}+2(3g^{2}+{g'}^{2})g^{2}\frac{(s_{\psi}^{2}-c_{\psi}^{2})^{2}}
{s_{\psi}^{2}c_{\psi}^{2}}\right)\right. \nonumber\\
&+&\left.f^{2}\log\left(\frac{\Lambda^{2}}{m^{2}}\right)
(3g^{4}+{g'}^{4}+8g^{2}{g'}^{2})
-3f^{2}\frac{g^{4}}{1-\frac{M_{W'}^{2}}{\Lambda^{2}}}\right]\nonumber
\end{eqnarray}

\begin{eqnarray}
\lambda_{\phi^2}&=&\frac{8N_{c}}{(4\pi f)^{2}}(\lambda_{t}^{2}+\lambda_{T}^{2})\left(\Lambda^{2}-m_{T}^{2}
\log\left(\frac{\Lambda^{2}}{m_{T}^{2}}+1\right)\right)\nonumber \\
&+&\frac{3}{64 \pi^{2}f^2}\left[\frac{g^{2}}{c_{\psi}^{2}s_{\psi}^{2}}\Lambda^{2}-g^{2}M_{W'}^{2}\log\left(\frac{\Lambda^{2}}{M_{W'}^{2}}+1\right)\left(\frac{(s_{\psi}^{2}-c_{\phi}^{2})^{2}}{c_{\psi}^{2}s_{\psi}^{2}}-4 \right) \right]\nonumber \\
&+&\frac{3g^{'2}}{(4\pi f)^{2}}\Lambda^{2}\nonumber
\end{eqnarray}

\begin{eqnarray}
\lambda_{H^{2}\phi}&=&-\frac{4N_{c}}{(4\pi f)^{2}}\left[(\lambda_{t}^{2}+\lambda_{T}^{2})\Lambda^{2}-\lambda_{T}^{2}m_{T}^{2} \log\left(\frac{\Lambda^{2}}{m_{T}^{2}}+1\right)\right]\nonumber \\
&+&\frac{3g^{2}}{8(4f\pi)^{2}}\frac{s_{\psi}^{2}-c_{\psi}^{2}}{c_{\psi}^{2}s_{\psi}^{2}}\left(\Lambda^{2}-M_{W'}^{2}\log\left(\frac{\Lambda^{2}}{M_{W'}^{2}}+1\right)\right)\,,\nonumber
\end{eqnarray}
\vspace{0.5cm}

\textbf{b. Coefficients coming from effective operators}
\vspace{0.3cm}

\emph{Modelo I}
\vspace{0.3cm}
\begin{eqnarray}
\lambda_{fg}^{\rm EO}&=& \frac{a}{8}\left(\frac{g^2}{s_{\psi}^2c_{\psi}^2}+\frac{g^{'2}}{s_{\psi}^{'2}c_{\psi}^{'2}}\right)+2a'(\lambda_{t}^2+\lambda_{T}^2)\nonumber \\
{\lambda_{\phi^2}}^{\rm EO}&=&   \frac{a}{2}\left(\frac{g^2}{s_{\psi}^2c_{\psi}^2}+\frac{g^{'2}}{s_{\psi}^{'2}c_{\psi}^{'2}}\right)+8a'(\lambda_{t}^2+\lambda_{T}^2) \nonumber\\
{\lambda_{H^2\phi}}^{\rm EO}&=& \frac{a}{4}\left(g^2\frac{c_{\psi}^2-s_{\psi}^2}{s_{\psi}^2c_{\psi}^2}+g^{'2}\frac{c_{\psi}^{'2}-s_{\psi}^{'2}}{s_{\psi}^{'2}c_{\psi}^{'2}}\right)+4a'(\lambda_{t}^2+\lambda_{T}^2)\nonumber
\end{eqnarray}
\vspace{0.5cm}

\emph{Modelo II}
\vspace{0.3cm}
\begin{eqnarray}
\lambda_{fg}^{\rm EO} &=& \frac{a}{8}\left(\frac{g^2}{s_{\psi}^2c_{\psi}^2}\right)-\frac{a}{3}g^{'2}+2a'(\lambda_{t}^2+\lambda_{T}^2)\nonumber\\
{\lambda_{\phi^2}}^{\rm EO}&=& \frac{a}{2}\left(\frac{g^2}{s_{\psi}^2c_{\psi}^2}\right)+4a{g^{'2}}+8a'(\lambda_{t}^2+\lambda_{T}^2)\nonumber\\
{\lambda_{H^2\phi}}^{\rm EO}&=& \frac{a}{4}g^2\frac{c_{\psi}^2-s_{\psi}^2}{s_{\psi}^2c_{\psi}^2}+4a'(\lambda_{t}^2+\lambda_{T}^2)\nonumber\\
\mu^{2\,\rm EO}&=&a f^2 g^{'2}\nonumber
\end{eqnarray}

\end{document}